\documentclass[utf8]{FrontiersinHarvard} 

\usepackage{url,hyperref,lineno,microtype,subcaption}
\usepackage[onehalfspacing]{setspace}

\def\chandra{{\itshape Chandra\/}}

\def\hexp{\hbox{{\itshape HEX-P\/}}}

\def\hst{{\itshape HST\/}}

\def\rxte{{\itshape RXTE\/}}

\def\xmm{{\itshape XMM-Newton\/}}
\def\nustar{{\itshape NuSTAR\/}}
\def\xray{\hbox{X-ray}}
\def\etal{{et\,al.}}

\def\ltsima{$\; \buildrel < \over \sim \;$}
\def\simlt{\lower.5ex\hbox{\ltsima}}
\def\gtsima{$\; \buildrel > \over \sim \;$}
\def\simgt{\lower.5ex\hbox{\gtsima}}
\def\kms{\ifmmode{~{\rm km~s^{-1}}}\else{~km s$^{-1}$}\fi}
\def\lsim{\lower0.3em\hbox{$\,\buildrel <\over\sim\,$}}
\def\gsim{\lower0.3em\hbox{$\,\buildrel >\over\sim\,$}}
\def\msol{$M_\odot$}
\def\h2{H$_2$}
\def\flux{erg~cm$^{-2}$~s$^{-1}$}
\def\lum{erg~s$^{-1}$}

\def\arcsec{\mbox{$^{\prime\prime}$}}
\def\arcmin{\mbox{$^\prime$}}
\def\sfr{$M_{\odot}$~yr$^{-1}$}

\def\simput{{\ttfamily SIMPUT}}
\def\sixte{{\ttfamily SIXTE}}

\def\keyFont{\fontsize{8}{11}\helveticabold }
\def\firstAuthorLast{Lehmer {et~al.}} 

\def\Authors{Bret~D.~Lehmer,$^{1,2,*}$ Kristen Garofali,$^{3,12}$
Breanna A. Binder,$^{4}$  Francesca Fornasini,$^{5}$ Neven Vulic,$^{6}$ Andreas Zezas,$^{7,8,9}$ Ann Hornschemeier,$^{3}$ Margaret Lazzarini,$^{10}$ Hannah Moon,$^{11}$ Toni Venters,$^{3}$  Daniel Wik,$^{11}$ Mihoko Yukita,$^{12}$ Matteo Bachetti,$^{13}$ Javier A. Garc\'{i}a,$^{3, 14}$ Brian Grefenstette,$^{14}$ Kristin Madsen,$^{3}$ Kaya Mori,$^{15}$ and Daniel Stern$^{16}$ }
\def\Address{$^{1}$Department of Physics, University of Arkansas, 226 Physics Building, 825 West Dickson Street, Fayetteville, AR 72701, USA \\
$^{2}$Arkansas Center for Space and Planetary Sciences, University of Arkansas, 332 N. Arkansas Avenue, Fayetteville, AR 72701, USA \\
$^{3}$X-ray Astrophysics Laboratory, NASA Goddard Space Flight Center, Code 662, Greenbelt, MD 20771, USA\\
$^{4}$Department of Physics \& Astronomy, California State Polytechnic University, 3801 W. Temple Ave, Pomona, CA 91768, USA \\
$^{5}$Department of Physics and Astronomy, Stonehill College, 320 Washington Street, Easton, MA 02357, USA\\
$^{6}$Eureka Scientific, Inc., 2452 Delmer St., Suite 100, Oakland, CA 94602-3017, USA\\

$^{7}$Physics Department \& Institute of Theoretical \& Computational Physics, University of Crete, 71003 Heraklion, Crete, Greece\\
$^{8}$Harvard-Smithsonian Center for Astrophysics, 60 Garden Street, Cambridge, MA 02138, USA\\
$^{9}$Institute of Astrophysics, Foundation for Research and Technology-Hellas, 71110 Heraklion, Greece\\
$^{10}$Department of Physics \& Astronomy, California State University Los Angeles, Los Angeles, CA 90032, USA
\\
$^{11}$Department of Physics \& Astronomy, The University of Utah, 115 South 1400 East, Salt Lake City, UT 84112, USA\\
$^{12}$The Johns Hopkins University, Homewood Campus, Baltimore, MD 21218, USA \\
$^{13}$INAF-Osservatorio Astronomico di Cagliari, via della Scienza 5, I-09047 Selargius (CA), Italy\\
$^{14}$Cahill Center for Astronomy and Astrophysics, California Institute of Technology,  Pasadena, CA 91106, USA\\
$^{15}$Columbia Astrophysics Laboratory, Columbia University, New York, NY 10027, USA\\
$^{16}$Jet Propulsion Laboratory, California Institute of Technology, Pasadena, CA 91109, USA
}
\def\corrAuthor{Bret Lehmer}

\def\corrEmail{lehmer@uark.edu}

\begin{document}
\firstpage{1}
\onecolumn

\title{The High Energy X-ray Probe: Resolved X-ray Populations in Extragalactic Environments} 

\author[\firstAuthorLast ]{\Authors} 
\address{} 
\correspondance{} 

\extraAuth{}

\maketitle

\newpage

\begin{abstract}

\section{}

We construct simulated galaxy data sets based on the
{\it High Energy X-ray Probe} (\hexp) mission concept to demonstrate the significant advances in galaxy science that will be yielded by the \hexp\ observatory.  The combination of high spatial resolution imaging ($<$20~arcsec FWHM), broad spectral coverage (0.2--80~keV), and sensitivity superior to current facilities (e.g., \xmm\ and \nustar) will enable \hexp\ to detect hard (4--25~keV) X-ray emission from resolved point-source populations within $\sim$800 galaxies and integrated emission from $\sim$6000 galaxies out to 100~Mpc.  These galaxies cover wide ranges of galaxy types (e.g., normal, starburst, and passive galaxies) and properties (e.g., metallicities and star-formation histories).  In such galaxies, \hexp\ will: (1) provide unique information about
X-ray binary populations, including accretor demographics
(black hole and neutron stars), distributions of accretion states and state transition cadences; (2) place order-of-magnitude more stringent constraints on inverse
Compton emission associated with particle acceleration in starburst
environments; and (3) put into clear context the contributions from X-ray
emitting populations to both ionizing the surrounding interstellar medium in
low-metallicity galaxies and heating the intergalactic medium in the $z > 8$
Universe.

\tiny
 \keyFont{ \section{Keywords:} Early-type galaxies (429); Star formation (1569); Starburst galaxies (1570); X-ray binary stars (1811); X-ray astronomy (1810); Compact objects (288)}
\end{abstract}


\twocolumn

%
\section{Introduction}\label{sec:intro}
%

The quest to understand galaxy formation and evolution relies on a broad diversity of investigative tools that provide detailed insight into the variety of physical phenomena and source populations present within galaxies.  For example, galaxy star-formation histories are effectively probed through color-magnitude diagram (CMD) isochrone analyses \citep[e.g.,][]{Cho2016} and panchromatic spectral energy distribution fitting \citep[e.g.,][]{Con2013, Car2019, Lej2019}; galaxy interstellar medium (ISM) conditions of ionization and chemical abundances are constrained through UV-to-IR spectral line modeling \citep[][]{Kew2019}; and the cosmic history of star-formation and chemical enrichment among galaxy populations are probed by wide-and-deep multiwavelength extragalactic surveys \citep[][]{Bla2009,Mad2014}.

In the context of galaxy evolution, X-ray emission from normal galaxies, not dominated by active galactic nuclei (AGN), traces the most energetic of processes, providing critical information for a variety of phenomena. Supernovae and their remnants probe the detailed physics of exploding stars; diffuse emission from hot gas traces the impact of recent star formation on galactic and intergalactic scales and the depths of gravitational potential wells; and X-ray binaries (XRBs) provide unique constraints on compact object, binary, and massive-star population demographics, as well as the physics of accretion onto compact objects and its environmental impact.

For the last $>$20 years, \chandra\ and \xmm, in conjunction with several multiwavelength facilities, have enabled many investigations of the roles of X-ray emitting phenomena in galaxy populations that span a broad range of properties \citep[e.g., from starbursts to passive ellipticals; see][for recent reviews]{Gil2022,Nar2022}.  

Due to its sensitivity above 10~keV, the advent of \nustar\ provided unique insight into the nature of luminous XRBs detected in galaxies out to $\approx$10~Mpc that had not been possible with previous X-ray observatories.  In particular, relatively unobscured spectral signatures that distinguish accreting compact object types, black holes (BH) versus neutron stars (NS), in XRBs, and the state of accretion are most notable in the $E \approx$~4--30~keV spectral range \citep[e.g.,][]{Rem2006,Don2007}.  Accretion disk state transitions in BHs, boundary layer emission from accreting NSs, and accretion columns in pulsars all have distinguishable spectral shapes in this energy range and can be efficiently classified via intensity-color and color-color diagrams.

Some of the highlights of \nustar-based studies of galaxies include: (1) direct measurement of the ratio of BH to NS XRBs in star-forming galaxies, indicating a rise in the BH fraction above a luminosity corresponding to the Eddington limit for 1.4~$M_\odot$ \citep[][]{Vul2018}; (2) identification of enhanced fractions of accreting pulsars and BH high-mass XRBs (HMXBs) in star-forming environments \citep[e.g., the SMC, M33, NGC~253, and M83;][]{Wik2014,Yuk2016,Laz2019,Laz2023,Yan2022} versus weakly-magnetized $Z$-type/atoll NS low-mass XRBs (LMXBs) in passive environments \citep[e.g., M31 and M81;][]{Mac2016,Vul2018}; (3) first constraints on the level of inverse Compton emission in the starburst environment of NGC~253 \citep{Wik2014}; and (4) quantification of the growing dominance of ultraluminous X-ray sources (ULXs) in star-forming galaxy X-ray spectra with declining metallicity \citep[e.g., measured by the characteristic $E>7$~keV spectral turnover in ULXs][]{Leh2015,Gar2020}.  These studies have provided critical insights into galaxy formation and evolution: XRB population classifications and spectral constraints inform population synthesis models and the role of X-ray radiation and mechanical feedback on the interstellar and intergalactic mediums \citep[e.g.,][]{Sim2021,Her2022,Her2023,Mis2023}

While the advances from \nustar, as described above, have demonstrated the power of using hard X-rays to uniquely constrain properties of XRBs, such studies are limited to only low signal-to-noise detections of the brightest X-ray sources (typically $\simgt$~few~$\times 10^{38}$~\lum) in galaxies in the nearby Universe.  To make progress, we require the next generation hard X-ray observatory to be capable of addressing:
\begin{enumerate}

\item {\bf What are the distributions of accretion states and compact object types (BHs, NSs, pulsars) across the full diversity of galactic environments in the Universe (i.e., spanning broad ranges of morphological types, star-formation histories, and metallicity)?}  How does the diversity of ULX spectral types vary across these properties?  What does this tell us about the formation of the most massive stellar remnants in the Universe, the most extreme regimes of accretion, and the evolutionary pathways of LIGO/VIRGO gravitational wave sources?

\item {\bf What are the duty cycles of accretion state transitions in BH and NS XRB populations outside the Milky Way (MW)?}   How do these variability constraints inform the physics of accretion disks and corona across compact object and donor-star types?

\item {\bf What is the role of diffuse inverse Compton emission in starburst environments and do leptonic or hadronic particle accelerations drive these processes?}  What does this imply about mechanical feedback from supernovae?

\item {\bf How does star-formation history and metallicity impact the intrinsic and emergent broadband X-ray spectra of the source populations?}  How do X-ray emitting sources contribute to ionizing radiation in ISMs across galaxy types and the heating of the intergalactic medium in the $z \simgt 8$ Universe?
 
\end{enumerate}

To uncover the scientific potential of the {\it High Energy X-ray Probe} \citep[\hexp;][Madsen et~al.\ 2023]{Mad2018} for addressing the above questions, we perform detailed simulations for a variety of galaxies.  With our simulations, we demonstrate how \hexp\ will revolutionize studies of X-ray emission from normal galaxies. Additional companion papers have been submitted providing additional details for how \hexp\ will advance scientific insights from other X-ray emitting populations in galaxies, including, e.g., ULXs and AGN.

%
\section{\hexp\ Mission Design}\label{sec:mis}
%

The \hexp\ probe-class mission concept offers sensitive broad-band coverage (0.2--80\,keV) of the X-ray spectrum with exceptional spectral, timing and angular capabilities. It features two high-energy telescopes (HET) that focus hard X-rays, and soft X-ray coverage with a low-energy telescope (LET).

The LET consists of a segmented mirror assembly coated with Iron monocrystalline silicon that achieves a half power diameter (HPD) of 3.5\arcsec, and a low-energy DEPFET detector, of the same type as the Wide Field Imager \citep[WFI][]{Mei2020} onboard {\it Athena} \citep{Nan2013}. It has 512$\times$512~pixels that cover a field of view (FOV) of 11.3\arcmin$\times$11.3\arcmin. It has an effective passband of 0.2--25\,keV, and a full frame readout time of 2\,ms, which can be operated in a 128 and 64 channel window mode for higher count-rates to mitigate pile-up and faster readout. Pile-up effects remain below an acceptable limit of $\sim$1\% for a flux up to $\sim$100~mCrab (2--10 keV) in the smallest window configuration. Excising the core of the PSF, a common practice in X-ray astronomy, will allow for observations of brighter sources, with a  typical loss of up to $\sim$60\% of the total photon counts.

The HET consists of two co-aligned telescopes and detector modules. The optics are made of Ni-electroformed full shell mirror substrates, leveraging the heritage of \xmm\ \citep{Jan2001}, and coated with Pt/C and W/Si multilayers for an effective passband of 2--80\,keV. The high-energy detectors are of the same type as flown on \nustar\ \citep{Har2013}, and they consist of 16 CZT sensors per focal plane, tiled 4$\times$4, for a total of 128$\times$128 pixel spanning a FOV slightly larger than for the LET, of 13.4\arcmin$\times$13.4\arcmin.  The HET utilizes the same optics technology as \xmm\ and the PSF is energy dependent with an HPD of 10\arcsec\ at 3 keV, $\approx$17\arcsec\ at 20 keV, and increases at higher energies. For the purpose of simulations in this paper, an average HPD of 17\arcsec\ was used across the entire bandpass. 

For normal-galaxy studies, \hexp\ will resolve XRB populations within galaxies out to $D \sim$~30--50~Mpc and will constrain their broadband spectra to faint luminosity limits (typically $L < 10^{38}$~\lum).  For more distant galaxies, \hexp\ will constrain the broadband X-ray spectra for thousands of galaxies out to $\approx$100--200~Mpc.

%
%
\begin{figure*}
\centerline{
\includegraphics[width=18cm]{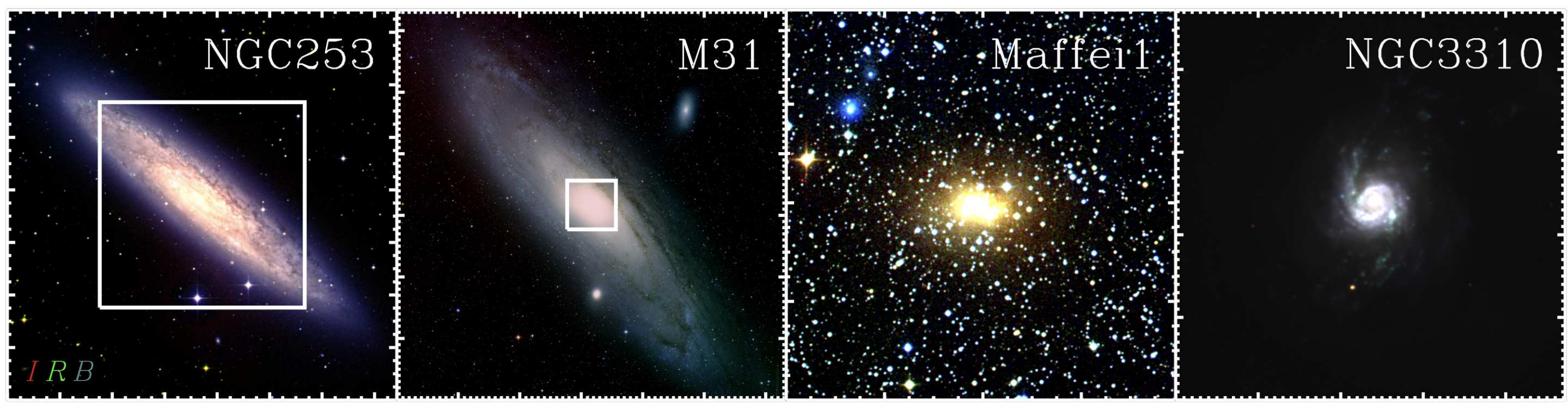}
}
\caption{
Optical images of the galaxy sample explored in this study.  Digitized sky survey images in $BRI$ (blue, green, red) are used for NGC~253, M31, and Maffei~1, with the exception of the $B$-band for NGC~253, which is from the Cerro Tololo Inter-american Observatory (CTIO).  For NGC~3310, Sloan Digital Sky Survey (SDSS) $gri$ bands are used.  The white square regions shown for NGC~253 and M31, as well as the entire FOV shown for Maffei~1, represent the 13\arcmin$\times$ 13\arcmin\ \hexp\ HET FOV that we simulate.  For NGC~3310, we show here the central 3.1\arcmin~$\times$~3.1\arcmin\ region of the HET FOV, which we use throughout the remainder of this paper.
}
\label{fig:sam}
\end{figure*}

\begin{table*}
\renewcommand\thetable{1}
{\footnotesize
\begin{center}
\caption{Galaxy Properties, \chandra\ Observation Log, and \hexp\ Simulation Summary}
\begin{tabular}{lccccccccccccc}
\hline\hline
 &  &  & & & \multicolumn{2}{c}{\chandra\ Observations}  & \multicolumn{6}{c}{\hexp\ Simulation}  \\
\multicolumn{1}{c}{Galaxy} & $D$$^{\star}$ &  SFR$^{\dagger}$ & $\log M_\star$$^{\dagger}$ & $Z$$^{\dagger}$ & $t_{\rm exp}$ & $\log(f_{\rm X}^{\rm min})$ & $t_{\rm exp}$ & \multicolumn{3}{c}{HET} & \multicolumn{2}{c}{LET} & $\log(L_{\rm 4-25~keV})$ \\
\multicolumn{1}{c}{Name} & (Mpc) & (\sfr) & ($M_\odot$) & ($Z_\odot$) & (ks) & (\flux) & (ks) & $N_B$ & $N_M$ & $N_H$ & $N_S$ & $N_M$ &  (\lum) \\
\multicolumn{1}{c}{(1)} & (2) & (3) & (4) & (5) & (6) & (7) & (8) & (9) & (10) & (11) & (12) & (13) & (14) \\
\hline\hline
NGC~253   & 3.94 & 5.6 & 10.8 & 1.0 & 156 & $-$16.3 & 500 & 81 &           60 &           50 &           67 &           61  & 36.8 \\
M31~bulge   & 0.776 & $<$0.3 & 10.5 & 1.6 & 676 & $-$16.3 & 50 & 81 &           64 &           49 &           73 &           67 & 36.0  \\
Maffei~1   & 3.4 & $\sim$0.0001 & 10.9 & 1.2 & 55 & $-$15.6 & 200 & 64 &           45 &           25 &           51 &           43  &  37.0 \\
NGC~3310   & 21.3 & 5.3 & 10.3 & 0.3 & 93 & $-$15.9 & 250 & 11 &           11 &            7 &           11 &           11  &  39.0 \\
\hline
\label{tab:obs}
\end{tabular}
\end{center}
Note.--- Col.(1): galaxy name.  Col.(2)--(5): distance, SFR, logarithm of the stellar mass, and metallicity estimate for each galaxy/region simulated.  Col.(6) and (7): archival \chandra\ observations cumulative exposure time and minimum 0.5--7~keV flux of the detected point sources. Col.(8): \hexp\ simulation exposure time. Col.(9)--(11): numbers of sources detected in our HET simulations in the broad ($B =$~4--25~keV), medium ($M =$~6--12~keV), and hard ($H =$~12--25~keV) bands. Col.(12) and (13) numbers of sources detected in the LET soft ($S =$~4--6~keV) and medium bands. Col.(14): \hexp\ HET simulation luminosity limit in the 4--25~keV band in units of \lum.\\
$^{\star}$Values of distance were extracted from the NASA Extragalactic Database (NED; http://ned.ipac.caltech.edu/).\\
$^{\dagger}$Listed values of SFR, $M_\star$, and metallicity were obtained from the literature.  For NGC~253 and NGC~3310 values of SFR and $M_\star$ were taken from the HECATE source catalog \citep{Kov2021}.  For the M31 bulge, the SFR was taken to be less than the total galaxy-wide SFR value of 0.3~\sfr\ \citep{Rah2016} and $M_\star$ was taken as $\approx$30\% of the total galaxy-wide stellar mass following the estimate in \citet{Tam2012}. For Maffei~1, the SFR was derived using the absorption-corrected H$\alpha$ luminosity from \citet{But2003} and the $L($H$\alpha$)--SFR relation from \citet{Ken1994} and $M_\star$ was taken from \citet{Mcc2014}.  Metallicities are based on gas-phase emission line measurements for NGC~253 \citep{Izo2006}, the M31 bulge \citep{San2012}, and NGC~3310 \citep{Eng2008}.  For Maffei~1, the metallicity is approximated following the stellar mass versus metallicity relation from \citet{Kew2008}.
}
\end{table*}
%

%
\section{Galaxy Sample and Example Science Program}\label{sec:gal}
%

For the purpose of demonstrating how \hexp\ will advance upon current constraints from existing \xray\ observatories (e.g., \nustar) and could help address the key questions posed in Section~\ref{sec:intro}, we perform \hexp\ simulations of a sample of galaxies that span a diversity of properties (see Fig.~\ref{fig:sam} and Table~\ref{tab:obs}).  Below, we describe the galaxy sample and the key scientific insights that can be afforded by studying these galaxies with \hexp.

{\it NGC~253:} Due to its proximity \citep[$D = 3.9$~Mpc;][]{Kar2004} and starburst nature (SFR/$M_\star \approx 10^{-10}$~yr$^{-1}$), NGC~253 represents an ideal target for investigating a variety of high-energy phenomena associated with star formation activity.  For example, a few dozen XRBs have been observed across the span of the galactic disk, probing the formation of populations of compact objects and binaries \citep[e.g.,][]{Pie2001}.  Diffuse thermal emission ($kT \approx 0.4$~keV) spans several arcminutes along the plane of the disk, with a hotter $\sim$1 keV gas component observed in a collimated kpc-scale outflow emanating
from the nuclear starburst \citep[e.g.,][]{Str2000,Str2002,Lop2023}. Finally, NGC~253 is one of only two starburst galaxies (the other being M82) that have been detected from hundreds of MeV to TeV energies, as a result of particle accelerations \citep[e.g.,][]{Abr2012,Hes2018}.

With $\approx$500~ks of cumulative \nustar\ exposure, only the most luminous \xray\ binaries could be detected above 10~keV.  These data indicated that ULX populations ($L > 10^{39}$~\lum) commonly exhibit high-energy ($E \simgt$~7 keV) spectral turnovers associated with super-Eddington accretion \citep[e.g.,][]{Pin2023} and the majority of HMXBs with $L \simgt 3 \times 10^{38}$~\lum\ (near the survey detection limit) have \xray\ colors consistent with either hard-state/intermediate-state BH XRBs or $Z$-type NSs, albeit with large uncertainties \citep[e.g.,][]{Wik2014}.  The data were further analyzed to place upper limits on the inverse Compton emission associated with starburst particle acceleration, in an attempt to distinguish between hadronic and leptonic particle dominance.

\nustar\ constraints on NGC~253 provided an enticing first demonstration of the scientific benefits afforded by focused hard X-ray imaging in a starburst environment not represented in the Local Group.  However, performing meaningful tests of theoretical binary-population models and particle acceleration models requires hard X-ray detection and compact object/accretion-state characterization of all X-ray binaries in outburst, as well as the disentanglement of hard X-ray emission from point-sources and diffuse emission.  \hexp\ has been designed with these goals in mind, and the high spatial resolution and high-energy capabilities will provide the needed constraints.

{\it The central bulge region of M31:} M31 is the nearest large spiral galaxy outside of the Milky Way.  At a distance of 776~kpc \citep[e.g.,][]{Dal2012}, M31 is close enough to resolve and detect populations of individual \xray\ sources at $>$10~keV with relatively short \nustar\ exposures (100~ks).  The bulge region of M31 hosts a high concentration of stars and LMXBs associated with the older stellar population of the galaxy.

\nustar\ constraints indicate that the X-ray sources in the bulge region are dominated by hard-state BHs and atoll/Z-source NSs, with the brightest source above 10~keV being an accreting pulsar \citep{Yuk2017}.  However, due to \nustar's relatively large PSF and the high density of sources, several bright sources in the nuclear region are undetected or have large photometric uncertainties due to source confusion.  An observatory like \hexp, which has both high spatial resolution and high-energy response, is needed to detect and classify the majority of the bright sources in the central bulge of M31 and monitor their accretion state transitions.

{\it Maffei~1:} Maffei~1 is one of the nearest \citep[$D \approx$~3--7~Mpc; e.g.,][]{Tik2018} massive elliptical galaxies ($M_\star \sim 10^{11}$~\msol), providing a unique target for studying a rich population of LMXBs in an old stellar environment in X-rays.  LMXB populations in elliptical galaxies have been observed to have steeply declining numbers of sources between $L \approx 10^{37}$~\lum\ and 10$^{39}$~\lum, with many of these bright sources crowded within small galactic footprints.  This, combined with the fact that there are few very luminous LMXBs (i.e., most have $L < 10^{38}$~\lum), has made it impossible to study such populations in hard X-rays with \nustar, which does not have the requisite angular resolution for resolving the populations.

A deep (177~ks) exposure with \chandra\ detected $\approx$150 \xray\ sources (spanning $L \approx$~$10^{36}$--$10^{38}$~\lum) within a 10\arcmin~$\times$~10\arcmin\ footprint centered on Maffei~1 (Ferrell et~al.\ in-prep).  \nustar\ imaging of the region reveals these sources as a few discrete concentrations of emission that trace the locations of the highest densities of sources.  The \nustar\ data constrain the average spectral properties of the population out to $\approx$20~keV.  However, to constrain the compact objects responsible for the accretion and the accretion states of this unique population requires hard X-ray data at high spatial resolution and sensitivity.

{\it NGC~3310:} NGC~3310 is a rare, yet nearby ($D \approx 20$~Mpc), low-metallicity galaxy ($\approx$0.3~$Z_\odot$) that has a substantial SFR ($\approx$8~\sfr).  Recent examinations of the metallicity-dependent formation rate of luminous HMXBs and ULXs in galaxies have shown an increase in the numbers of these sources with decreasing metallicity \citep{Map2010,Kov2020,Leh2021}.
While population synthesis models predict such trends, the key drivers are not well understood and may have multiple causes, including, e.g., the presence of more massive compact object remnants and weaker binary-destroying supernova kicks for low-metallicity stars, as well as smaller orbital separations throughout the evolution of low-metallicity binary stars \citep[][]{Lin2010,Wik2019}.
However, due to the rarity of low-metallicity galaxies in the local Universe, it has not been possible to search for key spectral and timing signatures that provide clues on the nature of HMXB populations in low metallicity galaxies in comparison to ULXs in higher metallicity galaxies.

Broadband (0.3--50~keV) measurements are capable of (1) characterizing contributions from accretion disks and coronae \citep[e.g.,][]{Wal2018}, (2) sensitively constraining cyclotron absorption lines \citep[e.g.,][]{Bri2022}, and/or (3) detecting NS pulsations at hard X-rays \citep[e.g.,][]{Bac2014} (a more detailed discussion of ULX science with \hexp\ is discussed in Bachetti et al.\ 2023, in-prep.). However, due to the rarity of low-metallicity galaxies in the local Universe, it has not been possible to search for key spectral and timing signatures that could provide concrete clues into the nature of luminous HMXBs and ULXs formed in low-metallicity environments.

As a result of its relatively high SFR, NGC~3310 contains a rich population of 14 \chandra-detected ULXs \citep{Ana2019}, providing an excellent laboratory for studying the properties of luminous HMXBs and ULXs in a low-metallicity environment.  However, at present, only \chandra\ and \xmm\ are capable of observing and disentangling the \xray\ properties of the individual ULXs, which are mainly present in a 20~arcsec diameter circumnuclear star-forming ring.

To investigate the hard X-ray spectrum of the aggregate population of ULXs in NGC~3310, a 140~ks \nustar\ observation was conducted \citep{Leh2015}.  \chandra\ and \nustar\ data were combined to constrain the 0.5--30~keV spectrum for the galaxy.  The galaxy-wide spectrum showed a steepening slope at $E \simgt 7$~keV, characteristic of ULXs; however, it was not possible to investigate the ULX properties individually.  \hexp\ will alleviate this limitation by resolving the ULX population in NGC~3310 and other low-metallicity galaxies, providing first constraints on the nature of the sources themselves and the diversity of their spectral properties as of function of luminosity in a low-metallicity environment.

In Figure~\ref{fig:sam}, we show optical color images of the above galaxies and highlight the specific \hexp\ areal extents of the galaxies that we simulate.

%
\section{Construction of Simulated \hexp\ Data Sets}\label{sec:sim}
%

\subsection{SIXTE Procedure}

To simulate \hexp\ data sets for the galaxy sample defined in Section~\ref{sec:gal}, we made use of the SImulation of X-ray TElescopes software package \citep[hereafter, \sixte;][]{Dau2019}.\footnote{See \url{https://www.sternwarte.uni-erlangen.de/sixte/} for \sixte\ code accessibility and documentation.}  \sixte\ is a flexible Monte Carlo based simulator, designed to model realistic X-ray data sets for a variety of astrophysical sources.  The simulation procedure in \sixte\ begins with the definition of a simulation input (\simput) file that contains the detailed astrophysical source population characteristics, including source sky locations, fluxes, spectra, intensity distributions, and time-dependent intensity variations.

Once a \simput\ file is defined, \sixte\ uses information about the observatory along with user-specified characteristics of the observation (exposure times, pointing direction, etc.) to generate a photon list (events list) for the observation.   In this process, photons are generated using a Monte Carlo procedure that accounts for the astrophysical source spectral shape, the energy-dependent effective area (via an ancillary response file; ARF), the expected photon signal on the detector (via a redistribution matrix file; RMF), effects of vignetting, and the instrument PSF.   Background photons can also be supplied for both instrumental and astrophysical sources, and \sixte\ provides flexibility in how background is applied; such background sources are included in our work. As such, \sixte\ simulations include all realistic sources of statistical noise for both sources and backgrounds that are expected in observations, and no two runs of the simulation are identical.

%
\begin{figure*}
\centerline{
\includegraphics[width=18cm]{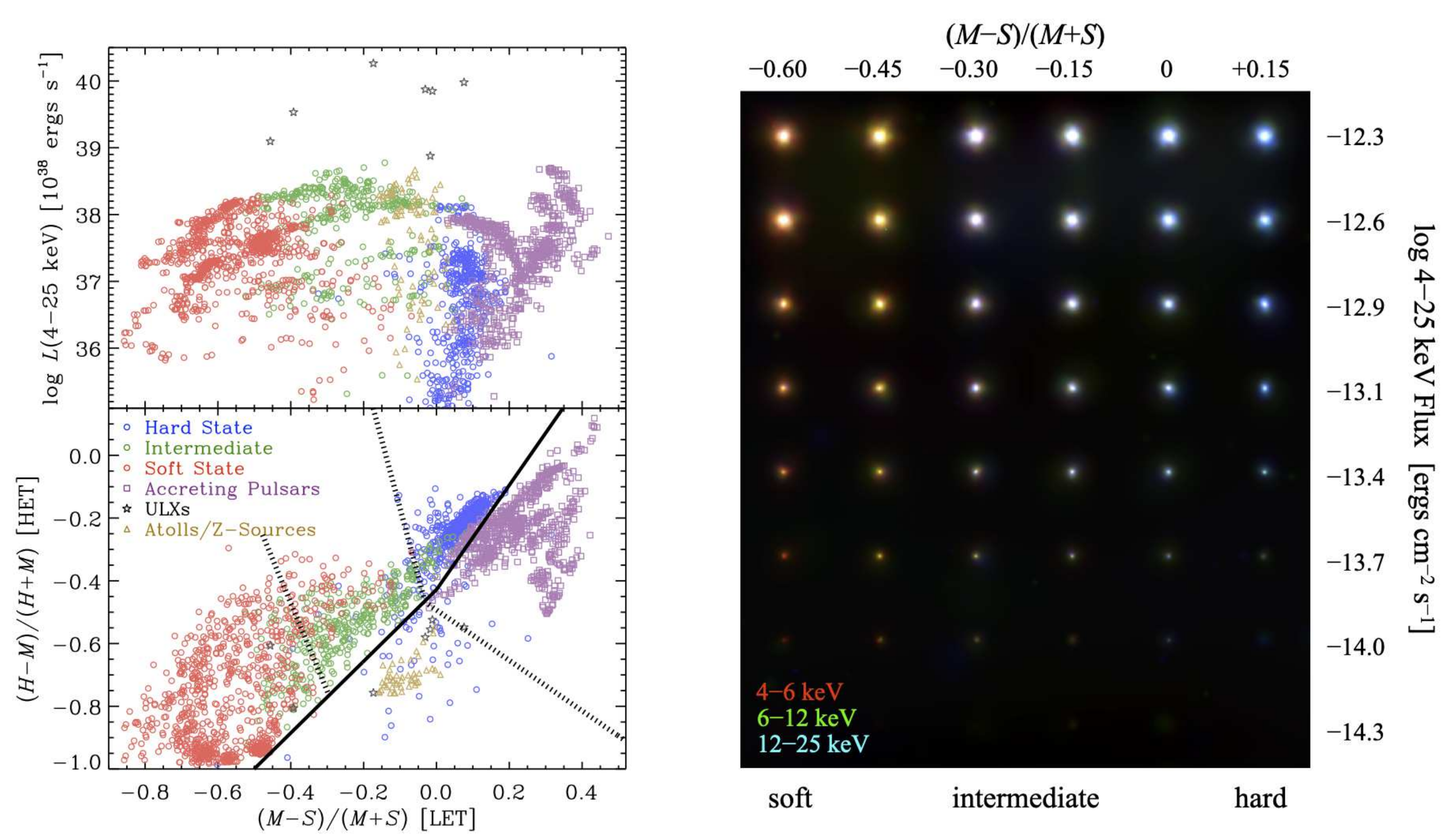}
}
\caption{
(Left) 4--25~keV luminosity versus \hexp\ LET medium-and-soft band color ({\it top\/}) and \hexp\ HET hard-and-medium band versus LET medium-and-soft band color ({\it bottom\/}) for a variety of Galactic XRBs sampled at several times by \rxte.  Bandpasses are defined as $S =$~4--6~keV and $M =$~6--12~keV, and $H =$~12--25~keV.  Various BH (hard, intermediate, soft states), NS (accreting pulsars, atolls/Z-sources), and ULX source populations have been color coded (see annotations).
(Right) 500~ks \sixte\ three-color \hexp\ HET simulation ($S$=red, $M$=green, $H$=blue) of BH XRB spectra with different values of $(M-S)/(M+S)$ [LET] color (see top annotation of columns) and 4--25~keV flux (see right annotation of rows).  The total number of 4--25~keV source counts is a weak function of color, however, this bandpass is most sensitive to detecting intermediate-state sources.
}
\label{fig:sens}
\end{figure*}

All the simulations presented below were produced using a set of response files that represent the observatory performance based on current best estimates as of Spring~2023 (see Madsen et~al.\ 2023). The effective area is derived from ray tracing calculations for the mirror design including obscuration by all known structures. The detector responses are based on simulations performed by the respective hardware groups, with an optical blocking filter for the LET and a Be window and thermal insulation for the HET. The LET background was derived from a GEANT4 simulation \citep{Era2021} of the WFI instrument, and the HET background was derived from a GEANT4 simulation of the \nustar\ instrument.  Both \hexp\ background simulations assume an L1 orbit.

\subsection{Sensitivity Limit Calculations and \hexp\ Exposure Calculations}\label{sec:sens}

The depths of our simulated observations are mainly driven by the goal of detecting XRBs and classifying their compact objects and accretion states in a variety of extragalactic environments.  To this end, our required exposures should be capable of both reaching interesting detection limits, while maintaining the ability to classify the X-ray sources. 

To accomplish this, we have constructed the luminosity-color and color-color diagnostic diagrams shown in the left panel of Figure~\ref{fig:sens}, which illustrate the \hexp\ diagnostic capabilities similar to those used for Galactic XRBs \citep[e.g., the ``q'' diagram of BH XRB accretion states][]{Bel2010}.  These diagrams were constructed using samples of XRBs in the Milky Way observed at multiple epochs with \rxte, as well as extragalactic ULXs studied by \nustar\ \citep[see, e.g., ][for further information]{Wik2014,Vul2018}.  The Milky Way samples include 2568 total \rxte\ observations of 6 BH XRBs, 9 accreting pulsars, and 11 weakly magnetized NS XRBs (i.e., Atoll/Z-sources), while the ULX samples include only single-epoch \nustar\ observations of 7 unique objects.  We note that our diagnostic diagrams are constructed from discrete observations of a finite number of bright sources that can be studied in detail, and that extragalactic populations are likely to span broader ranges of colors.  However, the physical variations that make these spectral classes distinguishable across the 4--25~keV spectral range are physical in nature, e.g., accretion disk and corona variations in BHs/NSs, boundary layer emission in NSs, and accretion columns in pulsars.  Thus, we expect that extragalactic XRBs will occupy similarly distinguishable regions of these color spaces.

The \rxte\ spectra at each epoch were fit using {\ttfamily xspec}, and best-fit models were convolved with \hexp\ responses to determine count-rates in $S =$~4--6~keV, $M =$~6--12~keV, and $H =$~12--25~keV bandpasses for both the LET and HET.  These bandpasses were chosen to optimize the color variation between compact objects and accretion states, while simultaneously guarding against the impact of absorption on X-ray spectra below $\sim$3~keV.  We note that the effective areas of the LET and HET are less than a factor of 2 different from each other across the $S$ and $M$ bandpass, with the HET effective area being larger than the LET for both bandpasses. These differences result in our simulated sources having average factors of $\approx$1.2 and 1.6 more counts for the HET versus the LET for the $S$ and $M$ bands, respectively.  However, due to the significantly sharper PSF of the LET compared to that of the HET, we find that the LET data were somewhat more sensitive than those of the HET.  However, the much larger effective area of the HET than the LET above 10~keV provided significantly more sensitive measurements for the $H$ band. Throughout the remainder of this paper, we therefore chose to adopt the LET color $(M-S)/(M+S)$ and HET color $(H-M)/(H+M)$ to illustrate the diagnostic power of \hexp.  More extensive analysis of \hexp\ data, however, can simultaneously utilize redundant LET and HET color diagnostics, and direct spectral fitting for brighter source populations for source classification.  

To determine the point-source detection limits as a function of accretion state (color) and flux, we first performed 500~ks \sixte\ simulations of BH XRBs that span the full range of BH XRB LET $(M-S)/(M+S)$ colors at a variety of input fluxes.  At the extremes of this range of colors are soft-state and hard-state BH XRBs, which show X-ray spectra that, respectively, drop and rise dramatically above  4~keV.  Thus, soft-state and hard-state BH XRBs will be brightest in the $S$ and $H$ bands, respectively, but both populations will be detectable in a broader 4--25~keV bandpass.

In the right panel of Figure~\ref{fig:sens}, we show a three-color ($S$ = red, $M$ = green, and $H$ = blue) \hexp\ HET image of our simulated sources at 500~ks depth, with each row representing a fixed 4--25~keV flux and each column representing a fixed spectral shape for BH XRBs (see annotations).  We chose a flux range of $5 \times 10^{-15}$ to $5 \times 10^{-13}$~\flux, which corresponds to a luminosity range of $10^{37}$--$10^{39}$~\lum\ at 4~Mpc (roughly the distance to NGC~253).  

To search for sources, we ran the {\ttfamily wavdetect} wavelet algorithm, available through the {\ttfamily CIAO} v.~4.14 software package,\footnote{https://cxc.cfa.harvard.edu/ciao/} on our 4--25~keV image.  Based on experimentation, we found that scales of 1.4, 2, 4, and 8 pixels and a false-positive probability threshold of $10^{-6}$ yielded reliable source lists and optimal numbers of real-source detections.  At 500~ks HET depth, all sources with 4--25~keV fluxes above $10^{-14}$~\flux\ ($\approx$2~$\times 10^{37}$~\lum\ at 4~Mpc) were detected, regardless of accretion state, and two intermediate-color sources with $5 \times 10^{-15}$~\flux\ ($10^{37}$~\lum\ at 4~Mpc) were also detected.  In these particular simulations, we did not detect any false positive sources (e.g., background fluctuations).  It is also clear from the three-color image in Figure~\ref{fig:sens} that the range of \xray\ accretion states are clearly discriminated by their $S$, $M$, and $H$ color down to the faintest source detections.

%
\begin{figure}
\centerline{
\includegraphics[width=8.5cm]{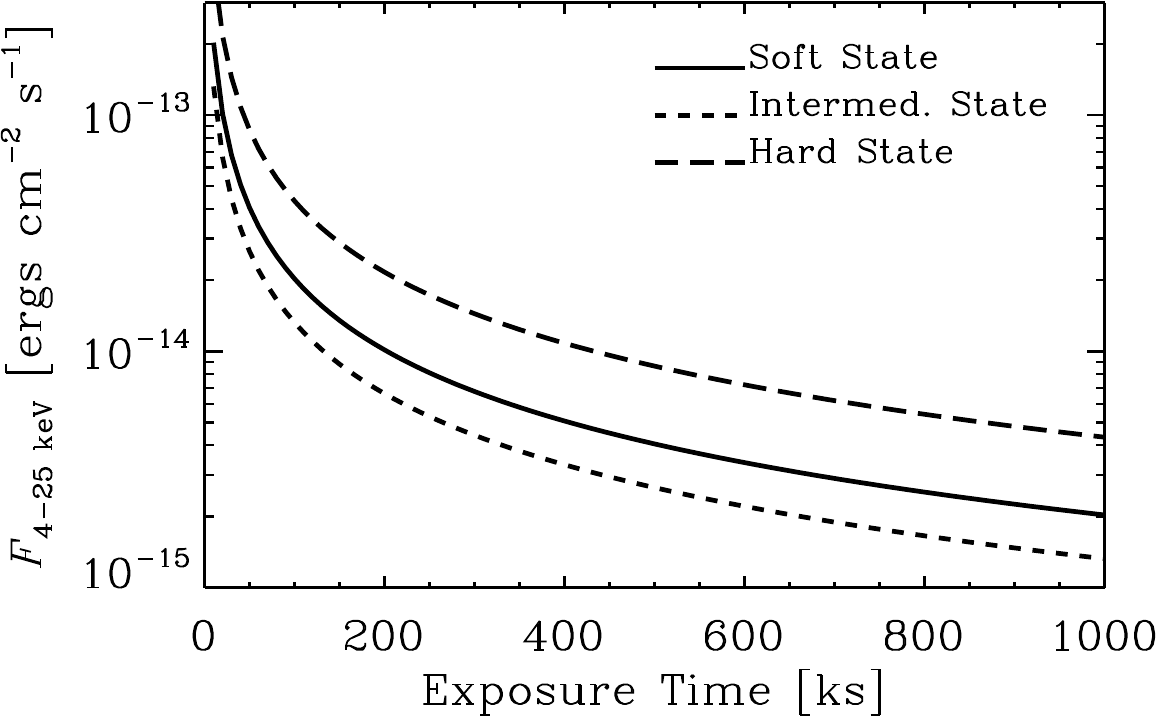}
}
\caption{
HET 4--25~keV flux detection limit as a function of exposure time for soft ({\it solid\/}), intermediate ({\it short-dashed\/}), and hard ({\it long-dashed\/}) state BH XRB spectral models.
}
\label{fig:flim}
\end{figure}

To clarify how the sensitivity limit depends on XRB accretion state and exposure time,  Figure~\ref{fig:flim} shows the approximate 3$\sigma$ 4--25~keV detection limit as a function of exposure time.  These curves are based on our HET sensitivity simulations (see Fig.~\ref{fig:sens}, right), in which we estimated 3$\sigma$ limits based on local count statistics (source and background) from 6~arcsec radii circular apertures for the sources given input 4--25~keV fluxes.  We found good agreement between these curves and {\ttfamily wavdetect} source detections from our simulations.

%
%
\begin{figure*}
\centerline{
\includegraphics[width=18cm]{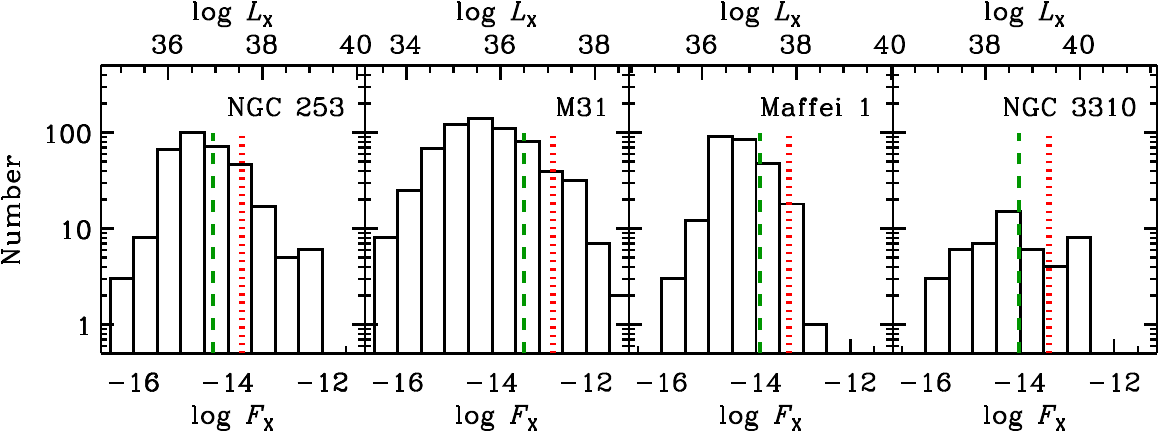}
}
\caption{
Distributions of input \chandra\ source catalog 0.5--8~keV fluxes (lower axis) and luminosities (upper axis) for each of the simulated galaxies in our sample.  The vertical lines show the approximate point-source 3$\sigma$ detection limits in our simulations for soft-state ({\it red dotted\/}) and hard-state ({\it green dashed\/}) BH XRB spectra and the simulation exposure times listed in Table~\ref{tab:obs}.  These limits indicate that even with the deepest simulated exposures, our input catalogs are likely to provide realistic accounting of X-ray source populations that include background contributions from sources below the individual source detection limits.
}
\label{fig:flx}
\end{figure*}

As shown in Figure~\ref{fig:flim}, the 4--25~keV HET sensitivity limit varies by a factor of $\approx$2 across accretion states and is most sensitive to sources in intermediate states, due to the combination of the spectral shape of this state and the shape of the effective area curve providing nearly optimal numbers of counts per unit flux.  Although we restrict our analysis in this paper to sources with 4--25~keV band detections, significant additional sensitivity can be gained by searching for sources in subband images and using these source positions to extract local photometry in adjacent bands. For example, the 2--6~keV LET could be used to more sensitively detect soft-state sources, and harder band photometry/limits could be used to constrain the compact object types and/or accretion states for such sources.

In the simulations that we discuss below, we have adopted exposure times that demonstrate \hexp's capability in addressing the key science goals outlined in $\S$\ref{sec:intro}.  These exposures are listed in Table~\ref{tab:obs}.  Specifically, NGC~253 and Maffei~1 bookend the range of stellar environments, i.e., starburst and passive environments, for which we can study how compact object and accretion state distributions vary across star-formation histories.  To characterize the properties of the bright XRBs in outburst requires that our observations reach 4--25~keV luminosity limits of $\approx$10$^{37}$~\lum.  In the case of NGC~253, we can also demonstrate the expected constraints on diffuse thermal and inverse Compton emission in starbursts.  To achieve these goals, we adopt exposures of 500~ks and 200~ks for NGC~253 and Maffei~1, respectively.

To monitor accretion state transitions among XRBs in extragalactic environments requires short, repeated observations that reach deep luminosity limits.  Given the proximity of a galaxy like M31 at $\approx$776~kpc, \hexp\ can reach a 4--25~keV luminosity limit of $\approx$10$^{36}$~\lum\ in $\approx$50~ks.  We therefore adopt this exposure in our simulations.  We note that several other galaxies (e.g., M33 and the Magellanic Clouds) are also near enough for fast and deep observations and are thus ideal for monitoring.

While low-metallicity galaxies are present in the nearby Universe (i.e., within $\approx$10~Mpc) and can be studied with \hexp, the majority of these objects are dwarf galaxies and contain few luminous XRBs and ULXs.  This makes it difficult to explore the role of metallicity in the properties of the ULX population.  NGC~3310 represents a more distant case ($D \approx$~20~Mpc) where multiple ULXs are observed in a single low-metallicity environment.  We adopt a 250~ks \hexp\ exposure to reach below the 4--25~keV detection limit of 10$^{39}$~\lum, corresponding to the typical definition of ULXs.  Our simulation of NGC~3310 serves as a benchmark for studying the effects of metallicity on ULX populations in galaxies out to $\approx$50~Mpc.

\subsection{Simulation Assumptions for our Galaxy Sample}

\subsubsection{Point-Source Properties}

As a starting point for each of our \hexp\ galaxy simulations, we utilized archival \chandra\ data with FOVs that overlap with existing \nustar-observed fields to define input point-source catalogs (sky locations and 0.5--7~keV fluxes and luminosities) for each galaxy.  We limited the regions of our analyses to 13\arcmin~$\times$~13\arcmin\ fields, the approximate fields of view for the existing \nustar\ observations and simulated \hexp\ footprints.  The regions were centered on the central coordinates of the galaxies, which we list in Table~\ref{tab:obs}. For all galaxies, except for M31, these regions cover the majority of the galactic footprints, as defined by ellipses that trace contours of constant $K_s \approx 20$~mag~arcsec$^{-2}$ galactic surface brightness \citep[see][for details]{Jar2003}.  However, for M31, the FOV is limited to \xray\ populations in the central bulge (see Fig.~\ref{fig:sam} for context), which accounts for a small fraction of the full galactic extent.

We identified archival \chandra\ ACIS observations that had aim-points within 5~arcmin of the galactic centers, and in the case of M31, we limited our search to observations with exposures $>$20~ks, to avoid very long processing times for the many monitoring observations available.  For all such ObsIds, we constructed merged data products and point-source catalogs following the procedures outlined in Section~3.2 of \citet{Leh2019}.  In Table~\ref{tab:obs}, we provide the \chandra\ observation log for the observations that were used to create the point-source catalogs.  The cumulative exposures of the \chandra\ observations are relatively long, $\approx$50--700~ks, which allows for very sensitive detections of point-source populations to faint limits.

%
\begin{figure}
\centerline{
\includegraphics[width=8.5cm]{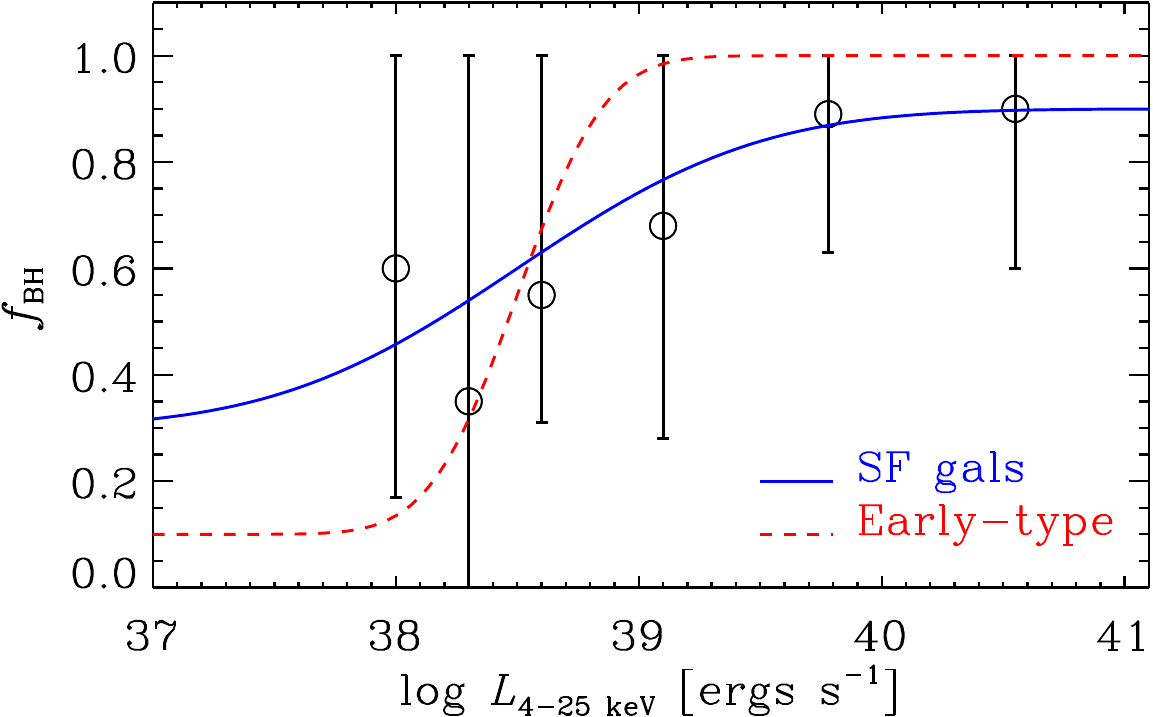}
}
\caption{
Current constraints on the BH fraction ($f_{\rm BH} = N_{\rm BH}/(N_{\rm BH} + N_{\rm NS})$) versus 4--25~keV luminosity for XRBs in star-forming galaxies studied by \citet{Vul2018} ({\it open circles with 1$\sigma$ error bars\/}).  Our heuristic models, which we adopt for the purposes of constructing our \sixte\ simulations, are displayed as a blue solid curve for star-forming galaxies (NGC~253 and NGC~3310) and a red dashed curve for passive environments (M31 bulge and Maffei~1).
}
\label{fig:fbh}
\end{figure}

In Figure~\ref{fig:flx}, we show the 0.5--8~keV flux and luminosity distributions for the \chandra\ point-source catalogs for each of the galaxies in our sample.  These point-source catalogs reach flux limits that are below the point-source detection limits of our simulated \hexp\ exposures.  As such, we consider our input catalogs to be sufficient for providing realistic simulated populations, including the fainter sources that contribute to the unresolved binaries and background AGN components, and we do not add additional contributions from \xray\ point sources that are not contained in the \chandra\ catalogs.

%
%
\begin{figure*}
\centerline{
\includegraphics[width=18cm]{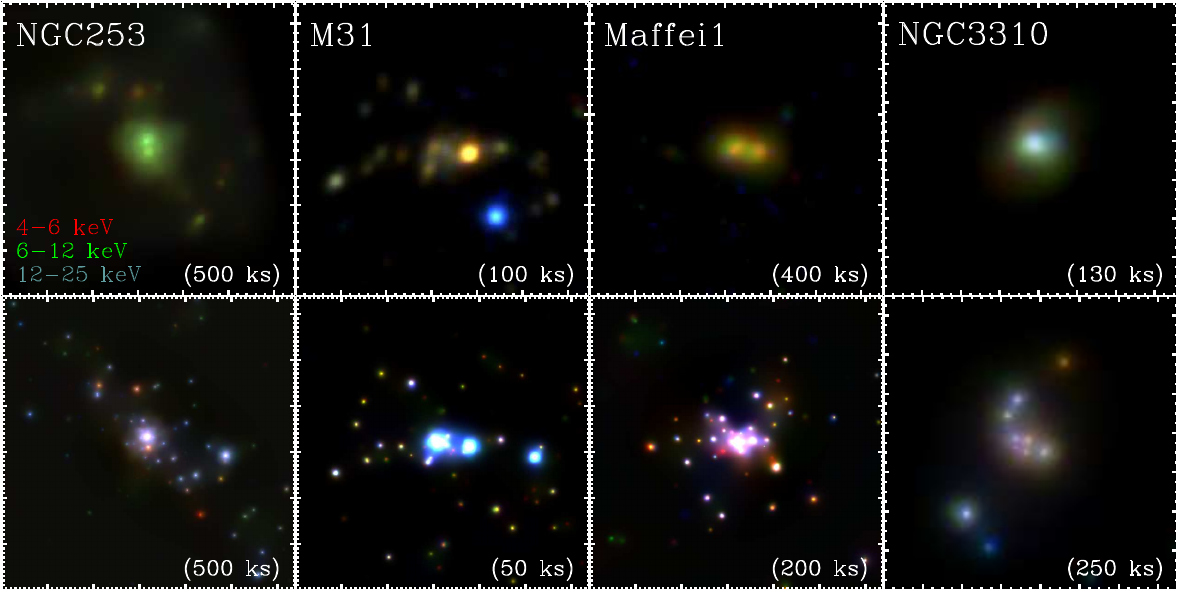}
}
\caption{
({\it Top Row\/}) \nustar\ three-color (red = 4--6~keV, green = 6--12~keV, blue = 12--25~keV) images of the galaxy sample outlined in Section~\ref{sec:gal}.  Each image has dimensions of 13\arcmin~$\times$~13\arcmin, except for NGC~3310, which has a 3.1\arcmin~$\times$~3.1\arcmin\ FOV.  These FOV correspond to the regions illustrated in the optical images of Figure~\ref{fig:sam}.
({\it Bottom Row\/}) Simulated three-color \hexp\ HET images of the same regions and colors as those displayed in the top panels.  Details of our simulations are described in Section~\ref{sec:sim}.  The improved PSF and throughput of \hexp\ provide dramatic gains in sensitivity over \nustar, resulting in major improvements in point-source characterizations with much larger numbers of X-ray sources detectable.  In addition to the HET data, the \hexp\ LET will simultaneously provide imaging across the 0.2--20~keV bandpass with 4$\times$ higher angular resolution than the HET.
}
\label{fig:gal}
\end{figure*}

To build our simulations in {\ttfamily SIXTE} requires not only the source positions and flux estimates, which we obtain from \chandra, but also spectral models that cover the full \hexp\ responses.  Current \chandra\ constraints on the spectral shapes of the \xray\ point sources are limited to the 0.5--8~keV band and have highly degenerate extrapolations to $E \simgt 8$~keV.  This is a primary driver of unique \hexp\ galaxy science.  Creating realistic spectral models therefore involves making highly subjective choices regarding the nature of the \xray\ sources.  We adopt an approach in which we utilize the 0.5--8~keV luminosity of a given source to select (1) a compact object type, either a BH or NS, and (2) an appropriate accretion state, given the compact object type.  The selection of luminosity, compact object type, and accretion state, along with some selection of galaxy-intrinsic absorption, uniquely defines a spectral model.  Below, we describe this procedure in detail.

To assign a spectral model for a given point source within our galaxy catalogs, we first selected a compact 
object type using a hypothetical BH fraction, $f_{\rm BH}$ (i.e., the fraction of all X-ray sources that have BH compact object accretors), versus luminosity diagram; we show such a diagram in Figure~\ref{fig:fbh}.  Recognizing that $f_{\rm BH}$ versus $L_{\rm X}$ is expected to vary with environment, we constructed two such curves, for star-forming and passive stellar host environments.  For the star-forming environment, we utilized a function consistent with \nustar\ constraints from \citet{Vul2018}, while for the passive environment, we used a function consistent with the BH fraction inferred by population synthesis model \xray\ luminosity functions (XLFs) produced for early-type galaxies NGC~3379 and NGC~4278 from \citet{Fra2009}.  We utilized the star-forming BH fractions for NGC~253 and NGC~3310 simulations and the passive environment BH fractions for the bulge of M31 and Maffei~1.  The compact object choice for a given source is selected by randomly drawing a number, $a$, from a uniform distribution in the domain of 0--1.  If  $a \le f_{\rm BH}$ ($a > f_{\rm BH}$), then we adopt a BH (NS) compact object type for the source.

With a compact object choice adopted for a given source, we used the diagrams shown in the left panels of Figure~\ref{fig:sens} to select an accretion state, which specifies a spectral model.  We accomplished this by first constructing probability distribution functions (PDFs) of \hexp\ LET $(M-S)/(M+S)$ color for various bins of 0.5--8~keV luminosity.  Such PDFs were created for BHs and NSs separately using the data shown in Figure~\ref{fig:sens}.  For BHs in all environments, the PDFs were constructed from the full distribution of MW BHs (soft, intermediate, and hard states) and ULX source spectra. For NSs in star-forming environments, we constructed the PDFs using the MW pulsar color distribution, while for NSs in passive environments, we used the MW Atolls/Z-sources to construst the PDFs.

Given a value of LET $(M-S)/(M+S)$ as drawn from the most relevant PDF, we identified a unique spectral template that best matches the drawn color.  These spectral templates contain spectral shapes that are characterized in {\ttfamily xspec} as the sum of accretion disk ({\ttfamily diskbb}) and Comptonization ({\ttfamily comptt}) models that have varying parameters (i.e., they span ranges of disk temperature, Comptonization temperatures and optical depths, and relative normalizations) and a fixed intrinsic absorption column density ($N_{\rm H} = 10^{21}$~cm$^{-2}$).  For a given simulated source, the best model is renormalized in \sixte\ to the observed 0.5--8~keV flux, as constrained by \chandra.

We note that in this paper, we perform the above selection procedure once for our simulations, thus resulting in a discrete choice of compact object types and accretion states among the sources.  Since this procedure is statistical, additional runs of the simulations will result in different draws of the numbers of sources in each category.  In practice, the resulting uncertainties on numbers of sources from this procedure will follow Poisson distributions, the uncertainties of which can be propagated into calculations of demographics (e.g., X-ray luminosity functions of various source types).

\subsubsection{Diffuse Emission in NGC 253}

For the case of NGC~253, we explicitly included diffuse emission in our simulations to explore the potential for \hexp\ to differentiate diffuse emission and point-source populations and also constrain the inverse Compton emission within the central starburst.  To model diffuse emission within \sixte\ requires an input model ``image'' that contains the intensity distribution projected onto the sky.  This input diffuse image can then be scaled to a user-chosen total flux with an \xray\ spectrum specified.  For NGC~253, we adopted separate diffuse emission images for the thermal emission that dominates at $E \simlt$~1--2~keV and the inverse Compton emission that is expected to be primarily confined within the central starburst region.

To construct the thermal diffuse emission map, we made use of the \chandra\ point-source catalog and data products for NGC~253.  We followed the \chandra\ X-ray center analysis thread for constructing a diffuse emission image after excluding known point sources.\footnote{\url{https://cxc.cfa.harvard.edu/ciao/threads/diffuse_emission/}}  The resulting point-source-free diffuse emission image was subsequently smoothed using {\ttfamily CIAO} task {\ttfamily csmooth} and sigma-clipped by setting map values that were $<1.5 \sigma$ below the median map value to zero.  The sigma clipping effectively removes pixels that are near the constant background value of the \chandra\ image and allows for the diffuse emission region to more effectively isolate true diffuse emission.

When modeling the thermal diffuse emission component spectral shape and galaxy-wide flux, we adopted the global constraints obtained by the \citet{Wik2014} \chandra-plus-\nustar\ investigation of NGC~253.  Specifically, we assumed three {\ttfamily apec} models with temperatures of $kT =$~0.2, 0.6, and 2~keV; hereafter, cool, warm, and hot components, respectively.   We assumed the warm and hot contributions were absorbed by column densities ({\ttfamily tbabs}) of $N_{\rm H} =$~1 and 7~$\times 10^{21}$~cm$^{-2}$ and the intrinsic model normalization ratios were $A_{\rm cold}/A_{\rm hot} = 0.13$ and $A_{\rm warm}/A_{\rm hot} = 0.4$.  When running our \sixte\ simulations, the total thermal model was normalized to a galaxy-wide 0.5--7~keV flux of $5 \times 10^{-12}$~\flux. 

For the inverse Compton component, we constructed a diffuse emission map following assumptions that were adopted in \citet{Wik2014}.  Specifically, we defined a relatively small elliptical region centered on the galactic center with dimensions $a = 30$\arcsec\ and $b = 8$\arcsec\ that was rotated 55~deg east-of-north to align with the galactic disk.  For the inverse Compton map, we chose to adopt a constant intensity across the elliptical region.  In \sixte\ we modeled the inverse Compton component using an absorbed power-law model with $N_{\rm H} = 3 \times 10^{21}$~cm$^{-2}$ and $\Gamma = 1.52$, with a 2--7~keV flux of $3 \times 10^{-13}$~\flux.

%
\section{Results}

\subsection{Creation of Data Products and Point-Source Catalogs}

Using the \simput\ catalogs and procedures discussed in $\S$\ref{sec:sim}, we ran \hexp\ LET and HET \sixte\ simulations for the galaxies in our sample using the exposure times discussed in $\S$\ref{sec:sens} and listed in Table~\ref{tab:obs}.  We created events lists and images in various bandpasses.  As our focus is on the characterization of point sources using intensity-color and color-color diagnostics, we chose to create images for both LET and HET in the broad $B =$~4--25~keV bandpass for source detection and $S$, $M$, and $H$ bands for compact object and accretion-state diagnostic purposes.  We also created LET 0.5--2~keV and 2--4~keV bandpasses and HET 25--50~keV and 50--80~keV bands.

In the top row of Figure~\ref{fig:gal}, we show three-color ($S$, $M$, and $H$ bands) \nustar\ images of our galaxy sample.  The FOVs of these images correspond to the regions indicated in the optical images shown in Figure~\ref{fig:sam}.  In the bottom panels of Figure~\ref{fig:gal}, we show corresponding three-color \hexp\ HET images in the $S$, $M$, and $H$ bands, the same bandpasses as shown by \nustar.  These simulated images alone provide a sense of the magnitude of science gained by \hexp\ over \nustar.  The sharper PSF and larger collecting area allow for substantially improved isolation of the point-source populations, enhanced sensitivity resulting in $\sim$1 order of magnitude more sources detected, and $\sim$1 order of magnitude more photons collected per source. 

%
%
\begin{figure}
\includegraphics[width=8cm]{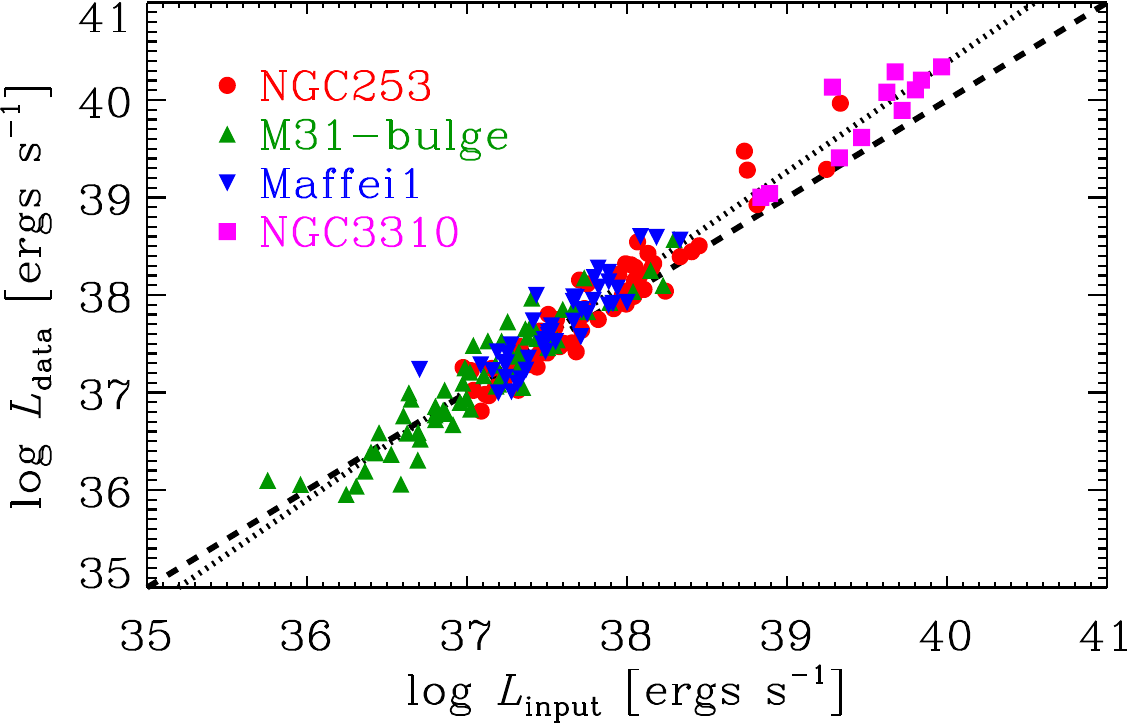} \\
\includegraphics[width=8cm]{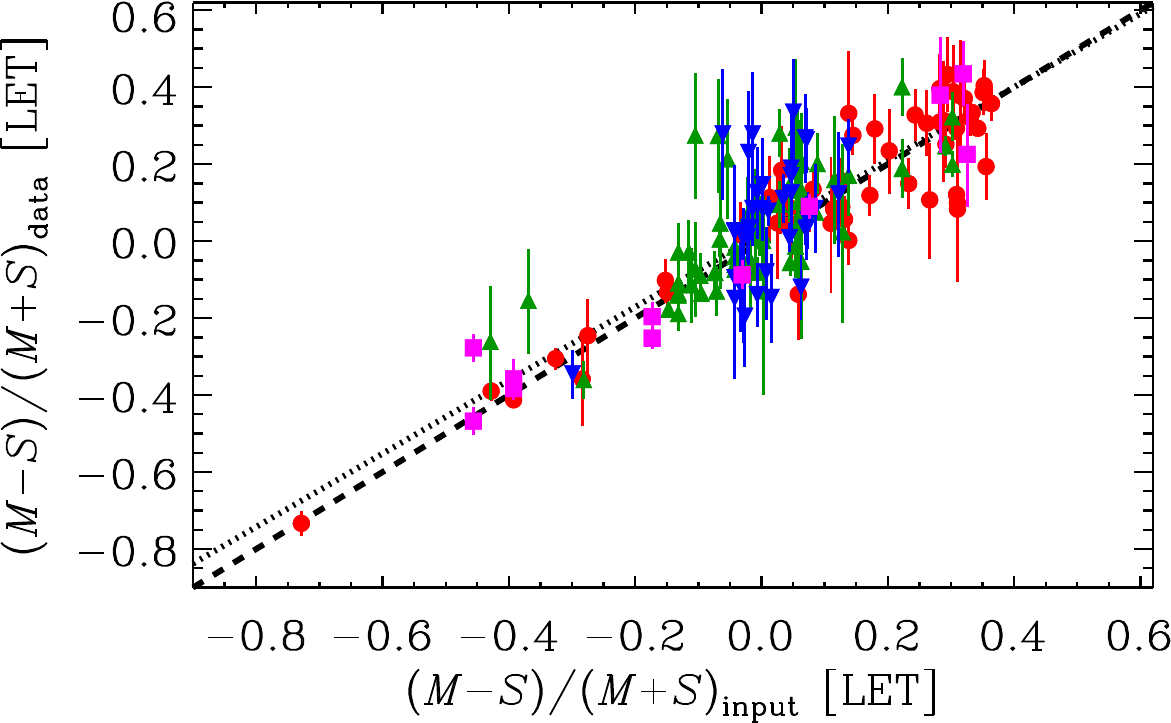} \\
\includegraphics[width=8cm]{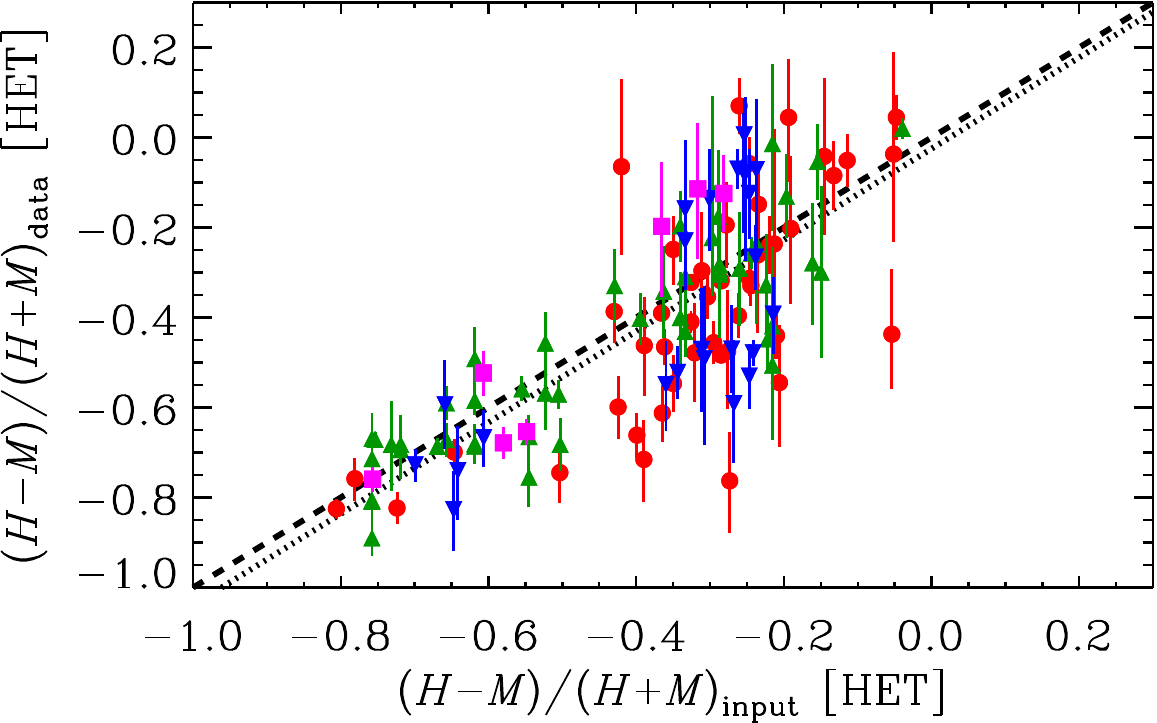}
\caption{
Assessments of the recovery of point-source properties from the four simulated galaxy data sets.  We show the recovery of a given point-source parameter from our analysis of the simulated images (ordinate) versus the values of the input parameters (abscissa).  In each panel, we show the one-to-one recovery line ({\it dashed black line\/}) and best-fit linear regression model ({\it dotted line\/}).  From top-to-bottom, we show the 4--25~keV luminosity, ($M-S$)/($M+S$) for the LET, and ($H-M$)/($H+M$) for the HET.  Most parameters are recovered well to within the uncertainties, with the exception of sources in crowded regions, where more detailed PSF fitting would be required.
}
\label{fig:cal}
\end{figure}

%
\begin{figure*}
\centerline{
\includegraphics[width=17cm]{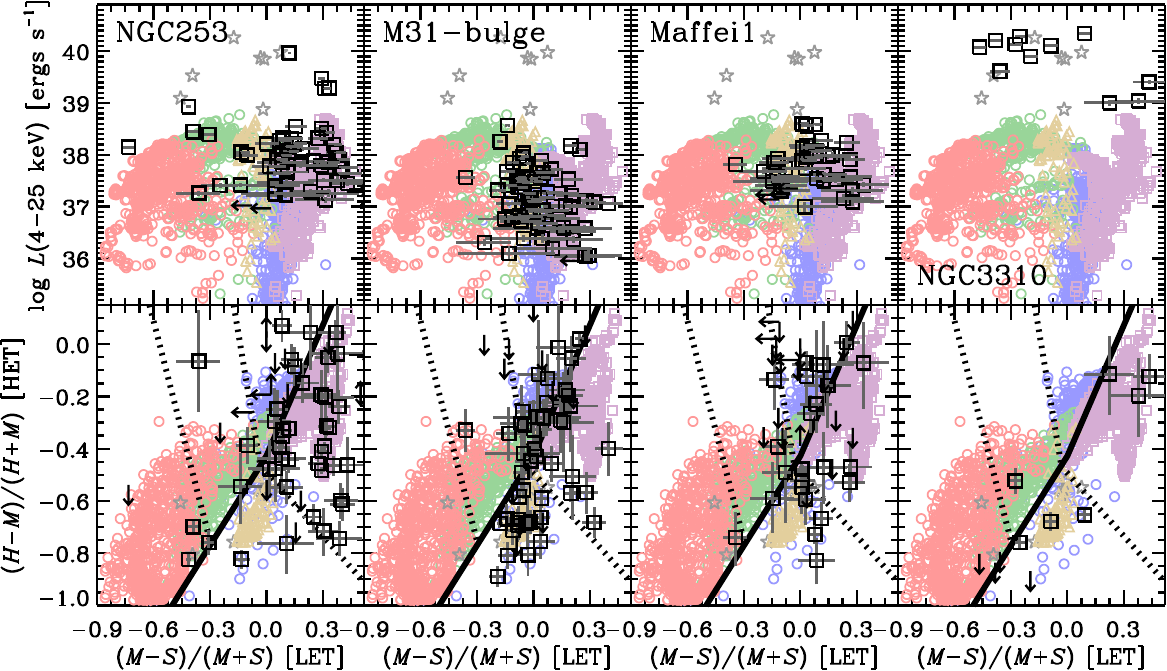}
}
\caption{
4--25~keV luminosity versus ($M-S$)/($M+S$) LET color ({\it top panels\/}) and $H-M$/($H+M$) HET color versus ($M-S$)/($M+S$) LET color ({\it bottom panels\/}) for each of our simulated galaxies.  Color symbols represent Milky Way and nearby ULX diagnostic locations and have the same meaning as they did in Figure~\ref{fig:sens}.  \hexp\ simulation constraints are shown as black open squares with 1$\sigma$ uncertainties or black arrows representing 3$\sigma$ upper limits.  In the bottom panel, lines represent divisions of classification that we adopt in this paper.
}
\label{fig:gcol}
\end{figure*}

For each simulated galaxy, we searched the $B$, $S$, $M$, and $H$ band images, for both the HET and LET, using the same procedure as employed in our calibration images: i.e., running {\ttfamily wavdetect} over scales of 1.4, 2, 4, and 8 pixels and a false-positive probability threshold of $10^{-6}$.  For simplicity, we created a catalog of sources based on the sources detected in the HET $B$-band images and cross-matched this source list to the $S$ and $M$ catalogs for the LET and the $M$, and $H$ band catalogs of the HET for the purposes of obtaining photometry for diagnostic classification purposes (see Fig.~\ref{fig:sens}).  When an HET $B$-band source was not detected in one of the subbands, 3$\sigma$ upper limits were calculated on the local source counts and propagated when calculating hardness ratio limits.  

For all bandpasses, we utilized source counts and background estimates from {\tt wavdetect} photometry, and applied aperture corrections to the photometry based on the areas used by {\ttfamily wavdetect} and our knowledge of the PSFs.  Typically, the photometry is extracted from regions that cover PSF fractions of $\approx$80--90\% for the LET and $\approx$40--60\% for the HET (approximately 1$\sigma$ ranges).  Uncertainties on the counts were based on Poisson estimates of the source and background count estimates, and these uncertainties are propagated to other calculations throughout the remainder of this paper.

In Table~\ref{tab:obs}, we list the total numbers of sources detected in each of the bands of interest.  The numbers of detected sources range from $\approx$10 for NGC~3310 to $\approx$80 for NGC~253 and the bulge of M31.  As we describe below, one of our key goals is to characterize XRB compact object types and accretion states in a variety of environments.  Accomplishing this requires accurate constraints on luminosities and colors that can enable such characterizations (see Fig.~\ref{fig:sens}).  In Figure~\ref{fig:cal}, we show how well we are able to recover 4--25~keV luminosity, ($M-S$)/($M+S$) color for the LET, and ($H-M$)/($H+M$) for the HET, given knowledge of the simulation input.  We assessed the accuracy of recovering these parameters by performing a basic linear regression for each set of measured and input values, finding best-fitting relation slopes of $1.12 \pm 0.02$, $0.95 \pm 0.05$, and $1.01 \pm 0.07$ for 4--25~keV luminosity, ($M-S$)/($M+S$)[LET], and ($H-M$)/($H+M$)[HET], respectively.  This indicates that the majority of the point-source properties are well-recovered in the galaxies and that \hexp\ has the angular resolution to accurately disentangle the spectral properties of source populations in a variety of extragalactic environments, even with simple analysis prescriptions.  We note, however, a subset of sources that lie in crowded regions (e.g., in NGC~3310 and the central regions of the other galaxies) have some offsets from the input parameters due to difficulty disentangling contributions from nearby sources, which we expect is driving the somewhat non-linear relation that we observe for the recovered 4--25~keV luminosity. For such cases, PSF fitting and source disentanglement procedures would need to be employed to obtain accurate photometry \cite[see, e.g.,][for the implementation of such procedures with \nustar]{Wik2014,Yuk2016,Vul2018,Yan2022}.  

In the sections below, we utilize our simulated data products and catalogs to demonstrate \hexp's power for addressing the scientific questions posed in $\S$~\ref{sec:intro}.

\subsection{A Census of XRB Compact Object Types and Accretion States in Diverse Extragalactic Environments}\label{sec:states}

In Figure~\ref{fig:gcol}, we show the intensity-color (4--25~keV luminosity versus ($M-S$)/($M+S)$ LET color) and color-color (($H-M$)/($H+M)$ [HET] versus ($M-S$)/($M+S)$ [LET]) diagnostic plots for each galaxy, including MW XRB and ULX comparisons in the background (see $\S$\ref{sec:sim} and Fig.~\ref{fig:sens} for details).  Without employing any quantitative assessments, we can see the signatures of our simulation assumptions coming through in these diagrams. For example, the star-forming galaxies, NGC~253 and NGC~3310, have source populations that occupy broad ranges of BH accretion states and contain substantial populations of pulsars.  In contrast, the early-type environments, i.e., the bulge of M31 and Maffei~1, have larger fractions of non-pulsating NSs and BHs that are consistent with being in hard states.

%
\begin{figure*}
\centerline{
\includegraphics[width=17cm]{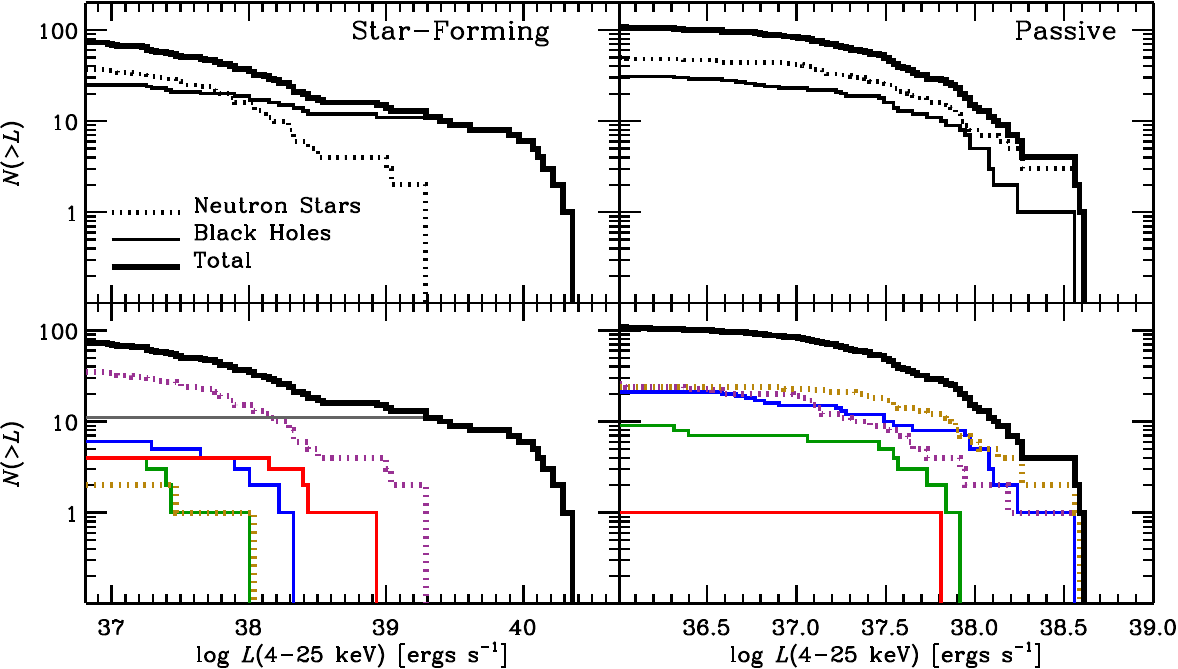}
}
\caption{
Observed 4--25 keV luminosity functions for star-forming ({\it left column}; NGC~253 and NGC~3310) and passive ({\it right column}; bulge of M31 and Maffei~1) environments from our simulations.  The luminosity functions have been decomposed into compact object types ({\it top panels\/}) and accretion state ({\it bottom panels\/}).  The accretion states include various BH states (thin solid lines), ULXs (gray), soft (red), intermediate (green), and hard (blue), while the NSs (dotted lines) are divided into weakly magnetize atoll/$Z$-sources (gold) and accreting pulsars (purple).  All classifications were based on the criteria defined in Section~\ref{sec:states} based on the simulated data.  The basic assumptions of our simulation inputs are recovered, including BH dominance for luminous sources in star-forming galaxies and NS dominance for XRBs in passive environments, and the dominance of pulsars and weakly magnetized NSs in star-forming and passive environments, respectively.
}
\label{fig:xlf}
\end{figure*}

While thorough probabilistic assessments would be required to properly classify sources as NSs and BHs, as well as characterize their accretion states, we can here provide first-order classifications of our sources based on their comparison with the MW and ULX comparison sample properties.  Following the methodology in \citet{Vul2018}, we divided the intensity-color and color-color spaces into discrete zones of occupation for various MW and ULX comparison sources and classify the \hexp\ simulated sources into compact object types and accretion states based on their zone location.  We optimized the locations of these zones to maximize the correct classifications of these zones.  For MW XRBs (non-ULXs), the following color division provides correct compact object type classifications for 98\% of the sources:
\begin{equation} \begin{split}
{\rm HR_2^{CO}} = & \left \{ \begin{array}{lr} -0.428 + 1.142 {\rm HR_1}  & ({\rm HR_1} < 0) \\ 
-0.428 + 1.667 {\rm HR_1},  & ({\rm HR_1} \ge 0) \\ 
\end{array}
  \right.
\end{split} \end{equation}
where ${\rm HR_1} = (M-S)/(M+S)$~[LET] and ${\rm HR_2} = (H-M)/(H+M)$~[HET].  Sources with HR$_2$ values above and below HR$_2^{\rm CO}$, as defined by equation~(1), are classified as BHs and NSs, respectively.  In Figures~\ref{fig:sens} and \ref{fig:gcol}, we show this dividing boundary as a solid line.

We further divide sources into accretion states using a similar approach.  We chose to divide sources into accretion states guided by the definitions illustrated in Figure~\ref{fig:sens}.  Specifically, we divided BH XRBs into soft, intermediate, and hard states, and NS XRBs into weakly magnetized atolls/$Z$-sources and accreting pulsars.  Again, we defined color boundaries that optimized the number of correct classifications.  For BHs, where HR$_2 >$~HR$_2^{\rm CO}$, the accretion-state boundaries were defined as:
\begin{equation}
{\rm HR_2^{soft/int}} =-1.593-2.804 {\rm HR_1}, 
\end{equation}
\begin{equation}
{\rm HR_2^{int/hard}} =-0.602-4.333 {\rm HR_1}, 
\end{equation}
where soft-state BHs have HR$_2 < {\rm HR_2^{soft/int}}$, intermediate-state BHs have ${\rm HR_2^{soft/int}} \le$~HR$_2 < {\rm HR_2^{int/hard}}$, and hard-state BHs have  HR$_2 > {\rm HR_2^{int/hard}}$.  For NSs, where HR$_2 \le$~HR$_2^{\rm CO}$, we used the equation,
\begin{equation}
{\rm HR_2^{mag}} =-0.493-0.803 {\rm HR_1}, 
\end{equation}
to divide between weakly-magnetized atoll/$Z$-sources (${\rm HR_2 }< {\rm HR_2^{mag}}$) and accreting pulsars (${\rm HR_2 } \ge {\rm HR_2^{mag}}$).

The above simple color selections reliably recover the correct accretion states for $\approx$90\% of the MW XRB populations shown in Figure~\ref{fig:sens}.  These boundaries are overlaid onto Figures~\ref{fig:sens} and \ref{fig:gcol} as dotted lines.

Using the above classification criteria and the measured constraints from our simulated data, as shown in Figure~\ref{fig:gcol}, we conducted basic classifications of the source populations based on either their measured colors or upper/lower limit constraints without detailed consideration of the uncertainties on these quantities.  With this simple approach, we found that $\approx$78\% of our sources had compact object classifications and $\approx$66\% had accretion state classifications that matched our input assumptions.  Sources with incorrect source classifications were primarily incorrectly classified due to photometric scatter, with some impact resulting from source crowding.  As such, the fraction of correct classifications significantly increases with increasing source counts.  The majority of the faint sources with incorrect classifications have uncertainties that overlap with the correct classification boundary.  Thus, more thorough analyses that account for the ambiguity of source classifications for faint sources and account for photometric uncertainties would be required to rigorously interpret these results.

Despite the above limitations, we can use our basic classifications to demonstrate the promise of using \hexp\ to infer information about XRBs in extragalactic environments.  In Figure~\ref{fig:xlf}, we show the 4--25~keV XLF for star-forming (NGC~253 and NGC~3310) and passive environments (bulge of M31 and Maffei~1) broken down into constituent compact object types and accretion states.  While the detailed shapes of these observed XLFs are highly influenced by variable imaging depth, incompleteness, and variations in galaxy properties (star-formation history and metallicity) that were used to construct them, we are able to recover unique new insights into the nature of the sources that contribute to the XLFs in line with the assumptions of our simulations.  

For instance, in star-forming galaxies, we can see the BH dominance among the bright-source populations and the pulsar dominance among the NS subpopulations.  For passive galaxies, we recover the NS dominance among LMXBs and show that the weakly-magnetized NSs dominate the NS population.  For both star-forming and passive galaxies, we see a steady rise in the fraction of hard-state BHs with decreasing luminosity, as the fraction of BHs in low-hard states outnumber soft/intermediate state sources.  We stress again that more thorough analyses that incorporate the impact of photometric errors and source crowding on classifications would be needed to concretely quantify and test population synthesis models.  Yet, our demonstration shows that these data will be extremely powerful for such tasks.

Of particular interest are the simulation results for the bulge region of M31.  These results show that \hexp\ will be capable of efficiently classifying XRBs in Local Group galaxies, even in crowded regions.  Our relatively short 50~ks simulated exposure reaches HET 4--25~keV luminosities of $\approx$$10^{36}$~\lum.  This demonstrates that \hexp\ could provide powerful monitoring of XRB populations in Local Group galaxies ($D \simlt 1$~Mpc), allowing for constraints on the duty cycles and the nature of accretion state transitions in XRB populations.

The Local Group includes galaxies representing diverse stellar populations, ranging from the actively star-forming Magellanic Clouds, dominated by stellar populations of $<$20~Myr and 30--70~Myr age for the SMC and the LMC, respectively \citep[e.g.,][]{SMCSFH,LMCSFH} to early type and late type spiral galaxies (M31, M33) and elliptical galaxies (M32). The availability of detailed star-formation history maps for these galaxies \citep[e.g.][]{SMCSFH,LMCSFH,M31SFH,M33SFH} from \hst\ data allow us to identify the X-ray counterparts of the X-ray sources, characterize their donor stars (e.g., supergiant, Be-XRB, giant, main-sequence donors), and associate them with individual star-formation events \citep[e.g.][]{Antoniou2010,Antoniou2016,Antoniou2019,Williams2013,Lazzarini2018,Lazzarini2021,Laz2023}.
Furthermore, these galaxies cover a wide range of metallicity, from 1/3~$Z_{\odot}$ for the SMC  \citep[e.g.][]{Antoniou2016} to solar metallicities for M31.  \hexp\ observations and monitoring of carefully selected fields, probing stellar populations of different ages \citep[c.f.][]{Antoniou2019},  would allow us to study directly the dependence of the compact object types and accretion states on the stellar population age, a key test for XRB population synthesis models.

\subsection{Spatial and Spectral Decomposition of X-ray Emitting Components in Nearby Galaxies}

In addition to \hexp's exquisite new capability for characterizing the nature of XRB populations using hard \xray\ colors, the high spatial resolution and broadband coverage of \hexp\ will work simultaneously to provide the best constraints to date on the broadband spectra of galaxies and the nature of the contributing components.  

In our simulation of NGC~253, we included contributions from XRBs, thermal diffuse emission, and inverse Compton emission (see Section~\ref{sec:sim}).  For this galaxy, the high spatial resolution enables the detection and isolated study of the point sources, allowing for the direct quantification of the spectral contribution that these sources make to the galaxy, which can be spatially separated from the remaining diffuse emission components (thermal and inverse Compton).  While this capability has been possible in the analysis of $\simlt$10~keV galaxy data from \chandra, and to some extent \xmm, \hexp\ will uniquely provide improvements in these constraints across the vast 0.2--80~keV spectral range without requiring coordinated observations of multiple facilities (as is often required with \nustar\ observations).  The LET will provide much improved spatial resolution over \xmm, and much larger effective area than \chandra, allowing for better spectral constraints on the the nature and absorption of point-sources (e.g., SNe and XRBs) and the thermal diffuse emission components in galaxies.  The $\simgt$10~keV improved sensitivity from the HET over \nustar\ will enable better constraints on the compact object population types and accretion states, as well as the contribution of the inverse Compton component of the ISM, which is predicted to grow in intensity over other \xray-emitting populations in this regime (see $\S$~\ref{sec:gal}).  

%
%
\begin{figure}
\centerline{
\includegraphics[width=8.5cm]{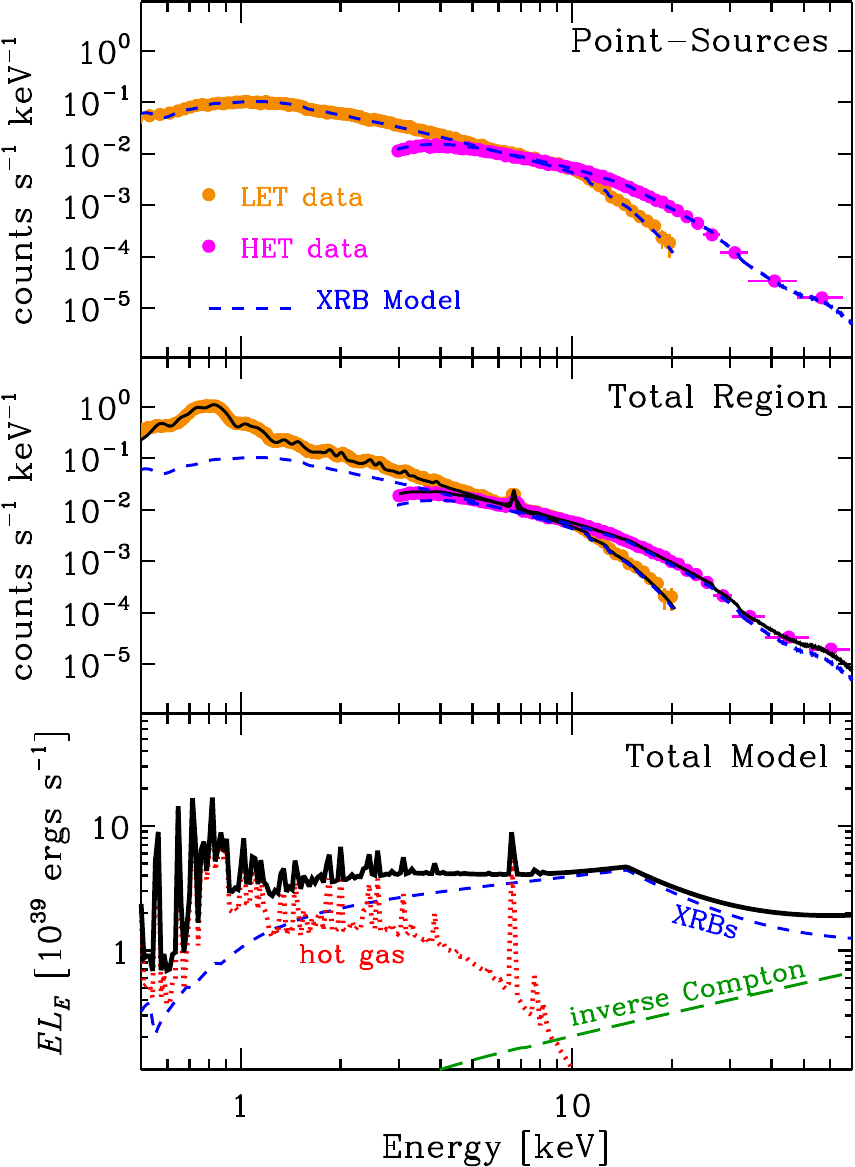}
}
\caption{
\hexp\ LET and HET spectra and best-fit models for the central 1\arcmin~$\times$1\arcmin\ starburst region of NGC~253.  Spectral fitting of the isolated point-source population data (bottom panel) can be used to constrain separately the XRB spectral contribution to the total spectrum (middle panel).  LET and HET data are shown as orange and magenta points with spectral binning and 1$\sigma$ uncertainties shown as error bars.  In the bottom and middle panels, the model contributions from XRBs are shown as blue dashed curves, while the remainder of the total spectrum (middle panel) is modeled due to diffuse thermal and inverse Compton emission.  Our unfolded best-fit model to the total spectrum is shown in the bottom panel, and includes contributions from thermal diffuse emission (red dotted), XRBs (blue short-dashed), and inverse Compton (green long-dashed).
}
\label{fig:n253}
\end{figure}

To demonstrate the analysis capabilities of \hexp, we investigated the simulated spectrum of the central 1\arcmin~$\times$~1\arcmin\ region, surrounding the galactic center where an ongoing starburst is present.  In the top panel of Figure~\ref{fig:n253}, we show the LET and HET spectral constraints as extracted from regions centered on the detected point-sources.  We performed spectral fitting of the point-source data using an absorbed broken power-law to account for ULXs, which have spectral steepening above $\approx$10~keV, plus a single power-law to account for the more typical low-luminosity XRB populations.  We show the best-fit ``XRB model'' in Figure~\ref{fig:n253} as a dashed blue curve.

In the middle panel of Figure~\ref{fig:n253}, we show the data extracted from the full 1\arcmin~$\times$~1\arcmin\ region, including the point-sources and diffuse emission.  We modeled the total spectrum using our XRB model, as constrained independently from the fit to the point-source data, plus a three-temperature thermal diffuse emission component and a single power-law with $\Gamma = 1.4$ to account for the inverse Compton emission.  For simplicity the three temperatures of the thermal component were modeled assuming $kT =$~0.2, 0.6, and 2.0~keV, with the normalizations on each of the components free to vary.  In this fitting procedure, we found that all components of the model (thermal emission, XRBs, and inverse Compton) were required in order to obtain a good fit to the data.  The solid curve in the middle panel of Figure~\ref{fig:n253} shows the best-fit overall model relative to the XRB contribution (blue dashed curve).  The bottom panel of Figure~\ref{fig:n253} shows the unfolded best-fit model in units of $E L_E$ ($\nu L_\nu$) with the model components overlaid.

While our constraints here are somewhat idealized by the assumptions of our modeling, the above procedure shows that \hexp\ will provide powerful new constraints on the spectral contributions from star-forming galaxies like NGC~253.

\subsection{Constraining the Broad X-ray Band Spectra of Galaxies for a Variety of Environments}

Thus far, we have discussed the scientific advances from \hexp\ observations of a few key nearby galaxies for addressing the scientific goals presented in $\S$~\ref{sec:intro}; however, definitively addressing these goals would require additional observations of galaxies in the nearby Universe.  To assess the reach of \hexp\ for studying galaxies, we utilized the Heraklion Extragalactic CATaloguE \citep[HECATE][]{Kov2021}, which contains an extensive compilation of the properties (e.g., distances, sizes, SFR, and $M_\star$) of 204,733 galaxies (based on the HyperLEDA catalog) out to $D \approx 200$~Mpc. In the discussion that follows, we assume normal galaxy parameters for our estimates of X-ray emission, but note that AGN activity will impact the X-ray emission from a non-negligible fraction of galaxies.  As discussed in other \hexp\ papers focused on AGN, \hexp\ will have excellent discriminating power for identifying AGN activity, even when heavily obscured or Compton thick.

Using the SFR and $M_\star$ values from the HECATE source catalog, we estimated the 4--25~keV luminosity of each galaxy using the scaling relation from \citet{Vul2018}:
\begin{equation}
       L_{\rm 4-25~keV}^{\rm scale} \, \approx  \\
       3.56 \times 10^{29}~M_\star + 1.90 \times 10^{39}~{\rm SFR},
\end{equation}
where $L_{\rm 4-25~keV}^{\rm scale}$ has units of \lum, $M_\star$ is in $M_\odot$ units, and SFR has units of \sfr.

%
\begin{figure*}
\centerline{
\includegraphics[width=16cm]{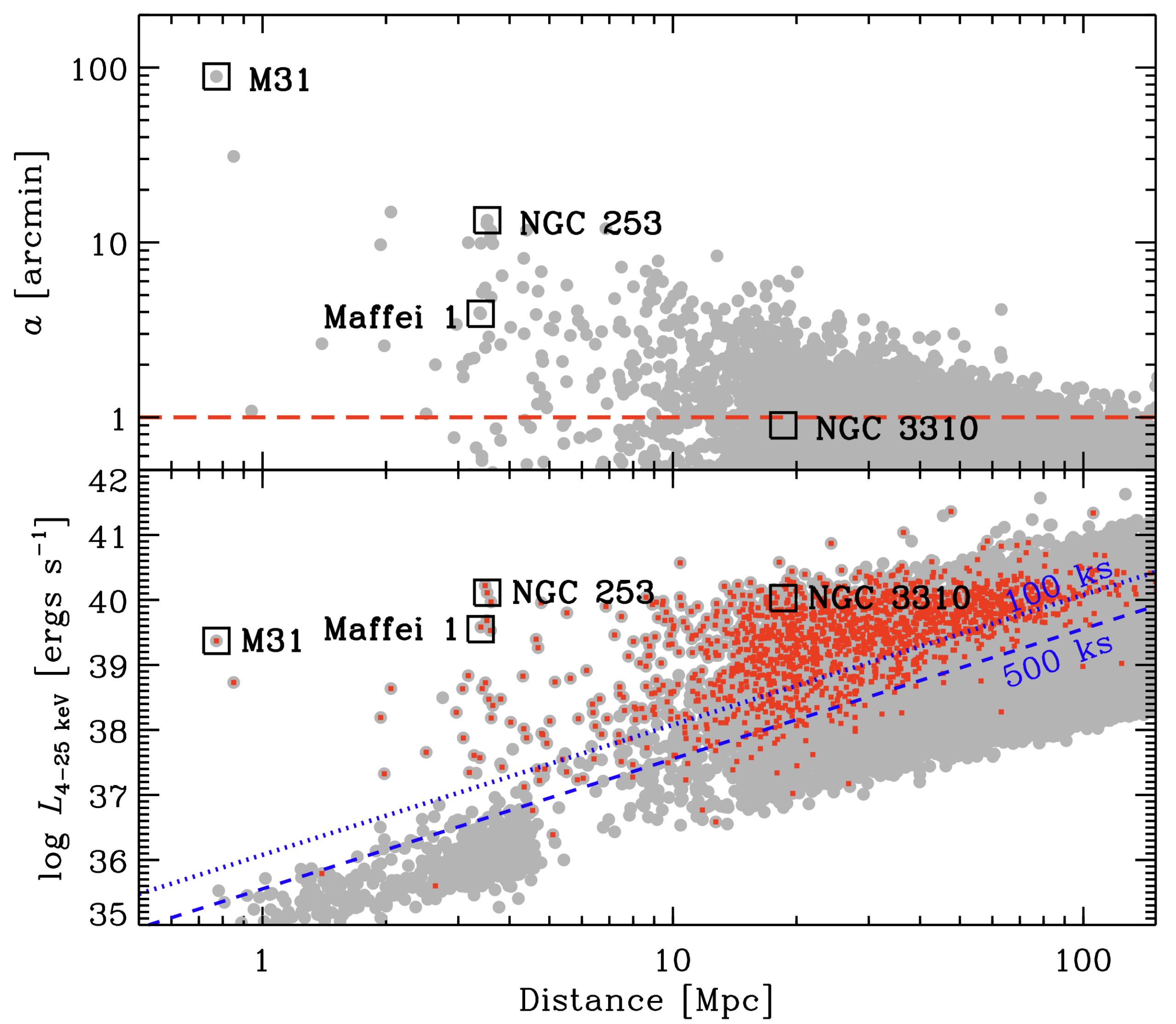}
}
\caption{
Estimated galactic semi-major axis (top panel) and scaling-relation estimated 4--25~keV luminosity (bottom panel) versus distance.  The underlying galaxy sample is from the HECATE source catalog \citep{Kov2021}, which is based on the HyperLEDA astronomical database \citep{Mak2014}  For reference, we show the locations of our simulated galaxies in both panels.  In the top panel, we indicate with a horizontal dashed red line the approximate limit above which \hexp\ can resolve galaxies into point-source populations. In the bottom panel, we show the expected 4--25 keV detection limits for the \hexp\ HET in 100~ks and 500~ks (blue lines) and indicate with small red squares galaxies with semi-major axes $\simgt$1~arcmin, in which populations could be resolved by \hexp.  At 100~ks depth, \hexp\ could study resolved populations in $>$1000 extended galaxies (major axes $\simgt$1~arcmin) and detect the \xray\ emission from 10,000s of galaxies in the local Universe.
}
\label{fig:hec}
\end{figure*}

In Figure~\ref{fig:hec}, we show the $D_{\rm 25}$ semi-major axis ($a$) and $L_{\rm 4-25~keV}^{\rm scale}$ versus distance for galaxies in the HECATE sample, along with locations of our simulated sample galaxies annotated.  From our simulation of NGC~3310, we showed that galaxies with extents smaller than $\approx$1~arcmin have XRB populations that are resolvable with \hexp\ HET.  From the top panel of Figure~\ref{fig:hec}, we see that there are many such galaxies in the $D \simlt 200$~Mpc Universe that could potentially be spatially resolved into their constituent point-source populations.  However, actually detecting such sources depends on the sensitivity of the observations and the density of sources.

In the bottom panel of Figure~\ref{fig:hec}, we show the point-source 4--25~keV detection limits achieved for the HET in 100~ks and 500~ks exposures.  We find that for the extended galaxies ($a \simgt 1$~arcmin), ULXs (i.e., $L > 10^{39}$~\lum) could be detected in $\simgt$800 galaxies at 100~ks depth, and XRBs with $L \le 10^{37}$~\lum\ could be studied in $\approx$50 (10) galaxies with 500~ks (100~ks) depth exposures.  In the quest to study the galaxy-integrated spectral properties of galaxy samples themselves, e.g., to explore how broadband \xray\ spectra vary as a function of galaxy properties, we estimate that $\approx$6000 galaxies could be detected at 100~ks depth, with $\simgt$300 such galaxies predicted to have luminosities $>$10 times higher than the 100~ks detection limit.

\begin{figure*}
\centerline{
\includegraphics[width=18cm]{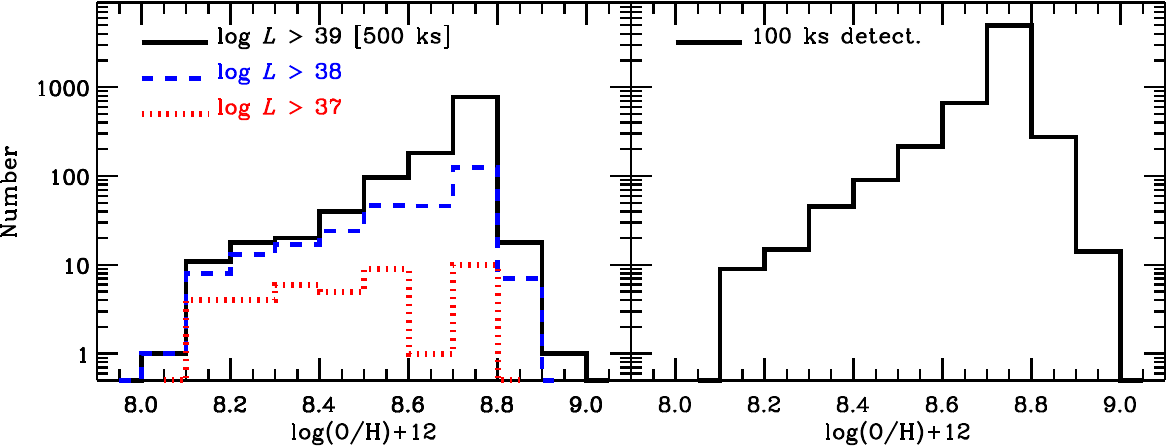}
}
\caption{
(Left) Distributions of metallicities for galaxies that could be detected and resolved into X-ray point source populations with 500 ks \hexp\ exposures down to limiting luminosities of $10^{37}$ (red dotted), $10^{38}$ (blue dashed), and $10^{39}$~\lum\ (black solid).
(Right) Distribution of metallicities for galaxies that would be detected in 100~ks \hexp\ exposures.  These histograms illustrate that many galaxies will be within reach for studying both resolved \xray\ populations and galaxy-integrated \xray\ emission with \hexp\ over a broad range of metallicity.
}
\label{fig:met}
\end{figure*}

To gain a sense of the diversity of metallicities that could be studied in these samples, we created Figure~\ref{fig:met}.  In the left panel, we show the distributions of galaxy metallicities for galaxies that could be detected to various luminosity limits in 500~ks and resolved into point source populations (i.e., with semi-major axes $a > 1$~arcmin).  In the right panel, we show the metallicity distribution of galaxies for which broadband constraints could be obtained in 100~ks depth exposures (right panel).  In this exercise, we utilized metallicity values in HECATE when available ($\approx$47\%) or estimates from the stellar mass versus metallicity relation from Table~2 of \citet{Kew2008}.  

Figure~\ref{fig:met} indicates that \hexp\ will be capable of studying resolved XRB populations down to $L \approx 10^{37}$~\lum\ and will obtain broadband SED constraints for galaxies spanning an order of magnitude in metallicity, spanning $Z \approx$~0.2--2~$Z_\odot$ (i.e., 12+$\log(O/H) \approx$~8--9).  Such a range represents the approximate mean metallicity of galaxy populations spanning redshifts $z \approx$~0--6, allowing for new interpretations for how the broadband X-ray spectra of galaxies, and their XRB constituents populations, evolved over the last $\simgt$90\% of cosmic history.  Constraints on the metallicity dependence of these populations will help inform population synthesis models, provide new constraints on the ionization properties of XRB populations in low-metallicity galaxies, and provide insight into how galaxies in the very early Universe (e.g., $z \simgt 8$) contribute to the heating of the intergalactic medium (see discussion and references in $\S$\ref{sec:intro}).

The fiduciary simulations and analysis presented in the previous sections demonstrates the ability of \hexp\ to revolutionize the fields of demographics of extragalactic XRBs by providing information on the compact object populations and accretion states for XRBs in a wide variety of galaxies.   Starting from our Local Group observations of selected regions in its four main galaxies (SMC, LMC, M31, and M33), probing stellar populations of different ages will provide the first constraints on the compact object populations and accretion state duty cycle for XRBs in different environments.  

%

In somewhat more distant galaxies (e.g., up to $\approx$10~Mpc), there are several spiral and dwarf galaxies covering broad ranges of star-formation history and metallicity, and many comprehensive panchromatic surveys of these galaxy samples have been conducted to provide subgalactic properties of the sources \citep[e.g.][]{Ken2003,11HUGS2008,ANGST2009,LEGUS2015,Lee2022,LVL.SEDs2023}.  Within the volume of  $D < 10$~Mpc, there are 7 galaxies with SFR above 5~\sfr\ and 15 galaxies with SFR between 1 and 5~\sfr. Based on the scaling relation of \cite{Leh2019},  we typically expect at least 10 to 20 sources with luminosity above $5\times10^{38}$\,\lum\ per galaxy (the 100~ks 4--25~keV detection limit at 10~Mpc), for the galaxies with SFR~$>5$~\sfr. Therefore, \hexp\ observations of even a few galaxies can provide a  statistically meaningful sample for which we will be able to study their compact object populations and accretion states.  Deeper observations to $\approx$$10^{37}$~\lum\  for a few of these galaxies, as we have shown here for NGC~253 and Maffei~1, would allow us to extend these studies to lower luminosity sources and perform comparisons with the Local Group galaxies.

Larger volumes include more extreme galaxies with even lower metallicities or higher SFR/$M_\star$ and SFR.  These are prime hosts of ULXs. In these galaxies, we can study in detail, with \hexp, the accretion state and compact object populations of sources accreting close to or above the Eddington limit. In addition, we will obtain detailed demographics of the ULX population at energies $\simgt$10~keV, which together with focused spectroscopic and timing studies of nearby bright ULXs (e.g., Bacchetti \etal\ 2023, in prep.), will provide a more clear picture of their nature across the broader universe.  

\section{Conclusions}

We have constructed simulations for the \hexp\ Probe concept of a sample of four nearby galaxies that span ranges of star-formation history and metallicity to demonstrate the power of using high spatial resolution broadband \xray\ data to constrain the nature of the underlying \xray\ emitting populations.  We have shown that the capabilities of \hexp\ are well designed for obtaining powerful new insights into these populations.  Specifically, we show that \hexp\ has the capabilities required to provide the following important constraints:

\begin{enumerate}

\item {\it HEX-P} will place informative broadband \xray\ constraints on large numbers of individual XRBs in galaxy populations that span broad ranges of morphology, star-formation history, and metallicity.  In particular, the sensitivity across the 4--25~keV range allows for compact object types (BHs and NSs) and accretion states to be discriminable, avoiding degeneracies in classification due to absorption.  With such data, it will be possible to investigate how the distributions of compact objects and accretion states vary as a function of formation age and metallicity, providing stringent new tests of stellar evolution including the impacts of binary evolution and close-binary interactions. 

\item The improved sensitivity of \hexp\ over \nustar\ will enable efficient detection and monitoring of the $\sim$100s of bright XRBs in Local Group galaxies to understand how the accretion states of the source populations transition as a function of time.  Such observations would allow for tests of accretion disk and corona modeling for a broadened range of binaries and mass-transfer scenarios.

\item The improved spatial resolution of the \hexp\ LET over \xmm, combined with the larger soft effective area over \chandra, will provide a new means for investigating thermal diffuse emission in galaxies without the contamination of bright XRBs.   \hexp\ HET constraints on the $\simgt$10~keV diffuse emission will provide powerful new constraints on the inverse Compton emission component that is predicted to be associated with particle accelerations in starburst galaxies.  The constraints on inverse Compton will aid in discriminating between hadronic and leptonic scenarios for particle accelerators in starbursts and will put into context the detected $\gamma$-ray emission from the nearby starburst galaxies NGC~253 and M82.

\item Taking a broader view, \hexp\ will be capable of constraining XRB population compact objects and accretion states for 100s of nearby galaxies ($D \simlt 50$~Mpc) and the broadband \xray\ spectra of thousands of galaxies out to $D \approx$~100--200~Mpc.  As such, \hexp\ can provide critical new constraints on populations of galaxies that span broad ranges of star-formation histories and metallicities, allowing for important new insights into how high-energy emitting sources emit throughout cosmic history.

\end{enumerate}

\section*{Author Contributions}

This paper is the result of work performed by the \hexp\ galaxies working group.  B.D.L. led the working group, led in the writing of all sections, and generated \hexp\ data simulations and figures.  K.G., B.A.B., F.F., N.V., and A.Z. provided additional text throughout the paper and detailed feedback on the manuscript.  J.G., B.G., Kristen~M., and D.S. provided text for Section~2. A.Z. provided simulated spectral parameters and corresponding accretion state information for Milky Way XRBs and ULXs that were used throughout Sections 4 and 5.  M.Z. and D.S. provided editorial comments on the manuscript.  All coauthors provided detailed input on the selection of galaxies for simulations, simulation setup and strategy, and analyses.  

\section*{Acknowledgments}
We are grateful to T. Dauser, M. Lorenz, and the SIXTE development team for their assistance with SIXTE simulations. B.D.L gratefully acknowledges support from the National Aeronautics Space Administration (NASA) Astrophysics Data Analysis Program (80NSSC20K0444).
N.V. acknowledges support for this work provided by the National Aeronautics and Space Administration through Chandra Award Number GO2-23065X issued by the Chandra X-ray Observatory Center, which is operated by the Smithsonian Astrophysical Observatory for and on behalf of NASA under contract NAS8-03060. K.G. was supported by an appointment to the NASA Postdoctoral Program at NASA Goddard Space Flight Center, administered by Oak Ridge Associated Universities under contract with NASA. The work of D.S. was carried out at the Jet Propulsion Laboratory, California Institute of Technology, under a contract with NASA. 

\bibliographystyle{Frontiers-Harvard} 
\bibliography{citations.bib}


\end{document}